\documentclass[12pt,english]{article}
\usepackage{lmodern}
\usepackage[T1]{fontenc}
\usepackage[latin9]{inputenc}
\usepackage{geometry}
\usepackage{amsmath}
\usepackage{amssymb}
\usepackage{graphicx}
\usepackage{esint}

\makeatletter
\usepackage{color}
\usepackage{hyperref}
\usepackage{cite}
\usepackage{babel}
\usepackage{mathdots}
\usepackage{amsfonts}
\usepackage{mathrsfs}
\usepackage{amsmath}
\usepackage{amssymb}
\usepackage{esint}
\usepackage{graphicx}
\usepackage{youngtab}
\usepackage{multicol}
\usepackage{multirow}
\usepackage{slashed}

\usepackage[titletoc]{appendix}

\allowdisplaybreaks

\numberwithin{equation}{section}

\date{}

\usepackage[usenames,dvipsnames,svgnames,table]{xcolor}
\hypersetup{colorlinks=true, citecolor=red, linkcolor=blue, urlcolor=violet}

\usepackage{titlesec}
\titleformat{\section}{\large\bf}{\thesection}{1em}{}
\titleformat{\subsection}{\normalsize\bf}{\thesubsection}{1em}{}

\makeatother

\usepackage{babel}
\begin{document}

\title{\textbf{\Large{Angular Momentum Generation from \\ Holographic Chern-Simons
Models\\\bigskip{}\bigskip{}}}}

\author{\textbf{\normalsize{Chaolun Wu}}}

\maketitle
\vspace{-16pt}

\begin{center}
\textit{Kadanoff Center for Theoretical Physics and Enrico Fermi Institute}
\par\end{center}

\begin{center}
\vspace{-35pt}

\par\end{center}

\begin{center}
\textit{University of Chicago, Chicago, Illinois 60637, USA}
\par\end{center}

\vspace{-10pt}

\begin{center}
\textit{\small{Email: }}\texttt{\small{chaolunwu@uchicago.edu}}
\par\end{center}{\small \par}

\begin{center}

\par\end{center}

\begin{center}
\thispagestyle{empty}
\par\end{center}
\begin{abstract}
We study parity-violating effects, particularly the generation of
angular momentum density and its relation to the parity-odd and dissipationless
transport coefficient Hall viscosity, in strongly-coupled quantum
fluid systems in 2+1 dimensions using holographic method. We employ
a class of 3+1-dimensional holographic models of Einstein-Maxwell
system with gauge and gravitational Chern-Simons terms coupled to
a dynamical scalar field. The scalar can condensate and break the
parity spontaneously. We find that when the scalar condensates, a
non-vanishing angular momentum density and an associated edge current
are generated, and they receive contributions from both gauge and
gravitational Chern-Simons terms. The angular momentum density does
not satisfy a membrane paradigm form because the vector mode fluctuations
from which it is calculated are effectively massive. On the other
hand, the emergence of Hall viscosity is a consequence of the gravitational
Chern-Simons term alone and it has membrane paradigm form. We present
both general analytic results and numeric results which take back-reactions
into account. The ratio between Hall viscosity and angular momentum
density resulting from the gravitational Chern-Simons term has in
general a deviation from the universal 1/2 value obtained from field
theory and condensed matter physics.
\end{abstract}
\newpage{}

{\hypersetup{linkcolor=black}

\tableofcontents{}

}

\bigskip{}

\bigskip{}
\bigskip{}

\section{Introduction}

When parity is broken, additional transport phenomena can take place
and reveal interesting underlying dynamical and topological structures
of the systems. Quantum Hall effect is a well known example. In fact,
in 2+1 dimensions, when parity is broken, in addition to Hall conductivity,
a few other parity-odd transport coefficients can also arise. These
transport coefficients had been systematically studied in \cite{Jensen:2011xb}
for relativistic fluids and recently in \cite{Kaminski:2013gca} for
non-relativistic fluids. Among them, Hall viscosity is the dissipationless
and parity-odd cousin of shear viscosity, just like Hall conductivity
can be viewed similarly compared to ordinary (longitudinal) conductivity.
The effect of Hall viscosity can be interpreted as a Lorentz-type
force (sometimes called the ``Lorentz shear force'') acting perpendicular
to the shear flow. Hall viscosity was first studied for various quantum
Hall states \cite{Avron:1995fg,Avron:1998,Tokatly:1,Tokatly:2,Read:2008rn,Haldane:2009ke,Read:2011}
and then for chiral superfluid states \cite{Read:2008rn,Read:2011}
and topological insulators \cite{Hughes:2011hv,Hughes:2012vg}. It
was also studied using general approaches such as linear response
theory \cite{Bradlyn:2012ea}, effective field theories \cite{Nicolis:2011ey,Hoyos:2011ez,Hoyos:2013eha,Son:2013rqa},
viscoelastic-electromagnetism \cite{Hidaka:2012rj} and quantum hydrodynamics
of vortex flow \cite{Wiegmann:1211,Wiegmann:1305,Wiegmann:1309}.
It was first noticed in \cite{Read:2008rn,Read:2011} and later re-derived
using more general methods in \cite{Nicolis:2011ey,Bradlyn:2012ea,Son:2013rqa}
that Hall viscosity is equal to a half of the orbital spin density
of the systems. In the absence of mechanical rotation or spin-orbit
coupling, this is the total angular momentum density of the system.
This reveals another interesting effect when parity is broken -- the
generation of angular momentum density and edge current. The microscopic
origin of such an angular momentum density varies for different systems,
but the common feature is the formation of vortices. For quantum Hall
states, this is from the cyclotron motion of the electrons or quasi-particles
in magnetic field. For chiral superfluids, this is due to the relative
orbital angular momentum of the two paired electrons in Cooper pairs
\cite{StoneRoy:2003,Sauls:2011,Tsutsumi:2012us}. The effect of the
non-vanishing angular momentum density and Hall viscosity is to accumulate
momentum and charges on the boundaries \cite{Wiegmann:1211} and to
induce an edge current.

Over the last decade, the gauge/gravity correspondence \cite{Maldacena:1997re,Gubser:1998bc,Witten:1998qj}
has often offered new insights to the understanding of strongly-coupled
quantum systems, such as quark-gluon plasma, superconductors, superfluids,
quantum Hall effects and topological insulators, just to name a few.
In this paper we are trying to understand the generation of angular
momentum density and its relation to Hall viscosity in 2+1-dimensional
strongly-coupled systems from the holographic point of view. In \cite{Saremi:2011ab}
a holographic model with dynamical gravitational Chern-Simons term
was first used to calculate Hall viscosity. This model was further
upgraded and numerically computed in \cite{Chen:2011fs,Chen:2012ti}.
Recently \cite{Cai:2012mg,Zou:2013fua} studied both Hall viscosity
and Curl viscosity using similar holographic models with Chern-Simons
terms. Spontaneous generation of angular momentum from holographic
models with gauge and gravitational Chern-Simons terms was also studied
in \cite{Liu:2012zm,Liu:2013cha}, and more recently in \cite{Liu:2014gto},
with focus on gapless systems. The common feature of all these studies
is that in their holographic actions, there are Chern-Simons terms
(gauge \cite{Wilczek:1987,Carroll:1989vb} or gravitational \cite{Jackiw:2003pm},
or both) coupled to a dynamical axion scalar field, which break parity
when the axion condensates. However, none of them reported to find
both Hall viscosity and angular momentum density at the same time,
thus does not yield a unified picture of them as that from the studies
using non-holographic approaches \cite{Read:2008rn,Read:2011,Hoyos:2011ez,Nicolis:2011ey,Son:2013rqa}.
Recently, \cite{Paper1} studied a different class of model -- the
holographic $p_{x}+ip_{y}$ model of \cite{Gubser:2008zu}. They found
both non-vanishing Hall viscosity and angular momentum density in
the superfluid phase, and showed that the ratio between Hall viscosity
and angular momentum density is a constant, at least near the critical
regime, and is numerically consistent with being $1/2$ in the probe
limit regime. This suggests an agreement with previous results from
\cite{Read:2008rn,Read:2011,Hoyos:2011ez,Nicolis:2011ey,Son:2013rqa}.
In fact, the holographic $p_{x}+ip_{y}$ model can be viewed as a
dual description to chiral superfluid states, like those studied in
\cite{StoneRoy:2003,Sauls:2011,Tsutsumi:2012us}, and it was from
computing Hall viscosity for such states (among others) that \cite{Read:2008rn}
first pointed out the relation between Hall viscosity and angular
momentum density. The holographic $p_{x}+ip_{y}$ model is different
from those Chern-Simons models in \cite{Saremi:2011ab,Chen:2011fs,Chen:2012ti,Cai:2012mg,Liu:2012zm}
that it does not contain Chern-Simons terms in the action, so the
action is perfectly parity-preserving. But the ground state breaks
spatial parity by locking it to non-Abelian gauge parity, which is
broken by the appearance of non-Abelian gauge connection, and this
is the only source for the emergence of Hall viscosity, angular momentum
density, and Hall conductivity (studied numerically earlier in \cite{Roberts:2008ns}).
Thus in holographic $p_{x}+ip_{y}$ model all the parity-odd transport
coefficients and angular momentum are generated in a unified way.

In this paper, we go back to the holographic gauge and gravitational
Chern-Simons models, and compute the angular momentum density using
the method proposed in \cite{Paper1}. It is worth noting that \cite{Liu:2013cha}
offers an alternative execution of the computation for angular momentum
density, and our general analytical results agree. We will show that
there is indeed a generation of angular momentum density accompanying
the emergence of Hall viscosity when the axion scalar condensates
spontaneously breaks parity. The angular momentum density receives
contributions from both the gauge and gravitational Chern-Simons terms.
It does not have a membrane paradigm form and part of its expression
is a bulk integral from the black hole horizon to the boundary. The
origin of this is that the vector mode fluctuations from which the
angular momentum density is calculated acquire effective masses through
mutual coupling in a non-trivial charged black hole background. On
contrary, the Hall viscosity has a membrane paradigm form because
the tensor mode fluctuations from which it is calculated remain massless.
The ratio between Hall viscosity and angular momentum density resulting
from the gravitational Chern-Simons term is not exactly a fixed constant,
but remains more or less unchanged as temperature is varied, except
at the very low temperature regime. The ratio depends on conformal
dimension of the condensate that breaks parity spontaneously.

The paper is organized as following. In section 2, we give the general
formalism of the holographic Chern-Simons models we are using and
the ground state ansatz. In section 3 and 4, we compute vector mode
and tensor mode fluctuations to obtain angular momentum density and
Hall viscosity, respectively. In section 5, we present numeric results
for the axion condensate phase, first in the probe limit and then
to include back-reactions. Conclusions and comments follow in section
6. Through out this paper, we will work in 3+1 spacetime dimensions.

\bigskip{}

\section{Holographic Chern-Simons Models}

For a general review on Chern-Simons modified gravity theory, we refer
readers to \cite{Alexander:2009tp}. In this section, we will only
list key ingredients relevant to the calculation of Hall viscosity
and angular momentum.

\subsection{Bulk and Boundary Actions}

The bulk action of our holographic Chern-Simons model in 3+1 dimension
is: 
\begin{equation}
S_{\textrm{bulk}}=\frac{1}{2\kappa^{2}}\int d^{4}x\sqrt{-g}\left\{ R-2\Lambda-\frac{1}{4}F_{\mu\nu}^{2}\right\} +S_{\vartheta}+S_{\textrm{CS}}+S_{\textrm{CS}}^{A}\:,
\end{equation}
where $F_{\mu\nu}=\partial_{\mu}A_{\nu}-\partial_{\nu}A_{\mu}$ is
the field strength of Maxwell field. The cosmological constant $\Lambda=-3/L^{2}$
and $L$ is the AdS radius. The real (pseudo) scalar $\vartheta$'s
action is 
\begin{equation}
S_{\vartheta}=\frac{1}{2\kappa^{2}}\int d^{4}x\sqrt{-g}\left\{ -\frac{1}{2}\left(\partial\vartheta\right){}^{2}-V\left[\vartheta\right]\right\} \:.
\end{equation}
We choose the potential to be 
\begin{equation}
V\left[\vartheta\right]=\frac{1}{2}m^{2}\vartheta^{2}+\frac{1}{4}c_{4}\vartheta^{4}\:,\label{AxionPotential}
\end{equation}
though in the actual calculation we will try to keep $V\left[\vartheta\right]$
general and not to implement this form until we have to. 

Abelian gauge Chern-Simons term is 
\begin{equation}
S_{\textrm{CS}}^{A}=\frac{1}{2\kappa^{2}}\int d^{4}x\sqrt{-g}\left\{ \frac{\lambda_{A}}{4}\Theta_{A}[\vartheta]\,^{*}\! FF\right\} \:,\label{GaugeCS_term}
\end{equation}
where $\lambda_{A}$ is the coupling constant, 
\begin{equation}
^{*}\! FF={}^{*}\! F^{\mu\nu}F_{\mu\nu}\:,
\end{equation}
and the dual field strength is 
\begin{equation}
^{*}\! F^{\mu\nu}=\frac{1}{2}\epsilon^{\mu\nu\alpha\beta}F_{\alpha\beta}\:.
\end{equation}
$\Theta_{A}[\vartheta]$ is a general functional of $\vartheta$. 

The gravitational Chern-Simons term is 
\begin{equation}
S_{\textrm{CS}}=\frac{1}{2\kappa^{2}}\int d^{4}x\sqrt{-g}\left\{ -\frac{\lambda}{4}\Theta[\vartheta]\,^{*}\! RR\right\} \:,\label{GravCS_term}
\end{equation}
where $\lambda$ is the coupling constant and the Pontryagin density
is defined as 
\begin{equation}
^{*}\! RR={}^{*}\! R^{\mu\nu\rho\sigma}R_{\nu\mu\rho\sigma}\:,
\end{equation}
and the dual Riemann tensor is 
\begin{equation}
^{*}\! R^{\mu\nu\rho\sigma}=\frac{1}{2}\epsilon^{\rho\sigma\eta\zeta}R_{\phantom{\mu\nu}\eta\zeta}^{\mu\nu}\:,
\end{equation}
where $\epsilon^{\rho\sigma\eta\zeta}$ is the Levi-Civita tensor.
We choose the convention%
\footnote{Our convention for $\epsilon^{\rho\sigma\eta\zeta}$ differs from
that of \cite{Saremi:2011ab,Paper1} by a sign, thus the corresponding
terms in both the Kubo formula and 2-point functions differ by a sign,
but the final expression for Hall viscosity remains the same because
these two signs cancel.%
} $\epsilon_{txyz}=\sqrt{-g}$. $\Theta[\vartheta]$ is a general functional
of $\vartheta$. Again we will try to keep its form general for as
long as possible in our calculation. A more detailed discussion of
its form will be presented in Section 5.

The boundary terms include the Gibbons-Hawking term

\begin{equation}
S_{\textrm{GH}}=\frac{1}{\kappa^{2}}\int_{\textrm{boundary}}d^{3}x\sqrt{-\gamma}K
\end{equation}
and a counter term

\begin{equation}
S_{\textrm{ct}}=-\frac{2}{\kappa^{2}R}\int_{\textrm{boundary}}d^{3}x\sqrt{-\gamma}\:,
\end{equation}
where $\hat{n}_{\mu}$ is the outgoing unit normal 1-form of the boundary
and $\gamma_{\mu\nu}=g_{\mu\nu}-\hat{n}_{\mu}\hat{n}_{\nu}$ is the
induced metric on the boundary. The extrinsic curvatures are $K_{\mu\nu}=\gamma_{\mu}^{\rho}\gamma_{\nu}^{\sigma}\nabla_{\rho}\hat{n}_{\sigma}$
and $K=K_{\mu}^{\mu}$. There is also a Chern-Simons boundary term,
analog to Gibbons-Hawking term, added such that the Dirichlet boundary
value problem is well posed: 
\begin{equation}
S_{\partial\textrm{CS}}=\frac{1}{2\kappa^{2}}\int_{\textrm{boundary}}d^{3}x\sqrt{-\gamma}\left\{ -\lambda\Theta[\vartheta]\hat{n}_{\rho}\epsilon^{\rho\sigma\gamma\delta}K_{\sigma}^{\phantom{\beta}\eta}\nabla_{\gamma}K_{\delta\eta}\right\} \:.
\end{equation}

\subsection{Perturbative Expansion of Bulk Action}

To compute 2-point functions, we perturbatively expand the on-shell
actions around the background up to second order in field fluctuations.
The background and fluctuations are 
\begin{eqnarray}
g_{\mu\nu} & = & \bar{g}_{\mu\nu}+h_{\mu\nu}\:,\nonumber \\
A_{\mu} & = & \bar{A}_{\mu}+a_{\mu}\:,\\
\vartheta & = & \bar{\theta}+\delta\theta\:,\nonumber 
\end{eqnarray}
where $\bar{g}_{\mu\nu}$, $\bar{A}_{\mu}$ and $\bar{\theta}$ are
the background and $h_{\mu\nu}$, $a_{\mu}$ and $\delta\theta$ are
fluctuations.%
\footnote{For axion, $\vartheta$ is the whole field, which equals background
plus fluctuation. $\bar{\theta}$ is the background part, which is
the same as $\theta(z)$ as defined in (\ref{Background}) to emphasize
the z-dependence. $\delta\theta$ is the fluctuation. $\theta_{0}$
and $\theta_{1}$ are the coefficients of non-normalizable and normalizable
modes of bulk background solution of the axion, as defined in (\ref{Axion_Asymptotic}).%
} To fully consider the back-reactions of the gauge fields on the metric,
we assume $h_{\mu\nu}$, $a_{\mu}$ and $\delta\theta$ are of the
same order. We also define the short-hand notations:
\begin{equation}
\Theta[\vartheta]=\bar{\Theta}+\delta\Theta\:,\qquad\textrm{where}\qquad\bar{\Theta}=\Theta[\bar{\theta}]\:,\quad\delta\Theta=\frac{\delta\Theta[\vartheta]}{\delta\vartheta}\Bigg|_{\vartheta=\bar{\theta}}\delta\theta\:,
\end{equation}
and similarly for $\Theta_{A}[\vartheta]$. The first order on-shell
action which is linear in fluctuations is 
\begin{eqnarray}
S_{\textrm{bulk}}^{(1)} & = & \frac{1}{2\kappa^{2}}\int d^{4}x\partial_{\mu}\Big\{\sqrt{-\bar{g}}\Big(\bar{\nabla}_{\nu}h^{\mu\nu}-\bar{\nabla}^{\mu}h-\bar{F}^{\mu\nu}a_{\nu}+\lambda_{A}\bar{\Theta}_{A}\,^{*}\!\bar{F}^{\mu\nu}a_{\nu}\nonumber \\
 &  & \qquad-\delta\theta\bar{\nabla}^{\mu}\bar{\theta}-\lambda\bar{\Theta}\,^{*}\!\bar{R}^{\nu\alpha\mu\beta}\bar{\nabla}_{\nu}h_{\alpha\beta}+\lambda h_{\alpha\beta}\,^{*}\!\bar{R}^{\mu\alpha\nu\beta}\bar{\nabla}_{\nu}\bar{\Theta}\Big)\Big\}\:.\label{S1_bulk}
\end{eqnarray}
The second order on-shell action quadratic in fluctuations is 
\begin{eqnarray}
S_{\textrm{bulk}}^{(2)} & = & \frac{1}{4\kappa^{2}}\int d^{4}x\partial_{\mu}\Big\{\sqrt{-\bar{g}}\Big[\frac{1}{2}h\bar{\nabla}_{\nu}h^{\mu\nu}+\frac{3}{2}h^{\mu\nu}\bar{\nabla}_{\nu}h-h^{\rho\sigma}\bar{\nabla}_{\rho}h_{\sigma}^{\mu}-2h^{\mu\rho}\bar{\nabla}^{\sigma}h_{\rho\sigma}\nonumber \\
 &  & \qquad+\frac{3}{2}h^{\rho\sigma}\bar{\nabla}^{\mu}h_{\rho\sigma}-\frac{1}{2}h\bar{\nabla}^{\mu}h-a_{\nu}\left(\frac{1}{2}\bar{F}^{\mu\nu}h+\bar{F}_{\rho}^{\phantom{\rho}[\mu}h^{\nu]\rho}+F^{(1)\mu\nu}\right)\nonumber \\
 &  & \qquad+\left(h^{\mu\nu}-\frac{1}{2}\bar{g}^{\mu\nu}h\right)\delta\theta\bar{\nabla}_{\nu}\bar{\theta}-\delta\theta\bar{\nabla}^{\mu}\delta\theta\Big]\Big\}\nonumber \\
 &  & +\frac{\lambda}{4\kappa^{2}}\int d^{4}x\partial_{\mu}\Big\{\sqrt{-\bar{g}}\Big[\frac{1}{2}\bar{\Theta}\,^{*}\!\bar{R}^{\nu\alpha\mu\beta}h_{\alpha}^{\sigma}\left(\bar{\nabla}_{\beta}h_{\sigma\nu}+\bar{\nabla}_{[\nu}h_{\sigma]\beta}\right)+\bar{\Theta}\,^{*}\!\bar{R}^{\nu\rho\mu\beta}h_{\rho}^{\alpha}\bar{\nabla}_{\nu}h_{\alpha\beta}\nonumber \\
 &  & \qquad-h_{\alpha\beta}h_{\rho}^{\alpha}\,^{*}\!\bar{R}^{\mu\rho\nu\beta}\bar{\nabla}_{\nu}\bar{\Theta}-\frac{1}{2}\bar{\Theta}\bar{\epsilon}^{\mu\beta\gamma\delta}\left(\bar{\nabla}_{\gamma}\bar{\nabla}_{\delta}h^{\nu\alpha}+\bar{\nabla}_{\gamma}\bar{\nabla}^{[\alpha}h_{\delta}^{\nu]}\right)\bar{\nabla}_{\nu}h_{\alpha\beta}\label{S2_bulk}\\
 &  & \qquad+\frac{1}{2}\bar{\epsilon}^{\nu\beta\gamma\delta}\left(\bar{\nabla}_{\nu}\bar{\Theta}\right)h_{\alpha\beta}\left(\bar{\nabla}_{\gamma}\bar{\nabla}_{\delta}h^{\mu\alpha}+\bar{\nabla}_{\gamma}\bar{\nabla}^{[\alpha}h_{\delta}^{\mu]}\right)-\,^{*}\!\bar{R}^{\nu\alpha\mu\beta}\delta\Theta\bar{\nabla}_{\nu}h_{\alpha\beta}\nonumber \\
 &  & \qquad+\,^{*}\!\bar{R}^{\mu\alpha\nu\beta}h_{\alpha\beta}\bar{\nabla}_{\nu}\delta\Theta\Big]\Big\}\nonumber \\
 &  & +\frac{\lambda_{A}}{2\kappa^{2}}\int d^{4}x\partial_{\mu}\Big\{\sqrt{-\bar{g}}\Big[\delta\Theta_{A}\,^{*}\!\bar{F}^{\mu\nu}a_{\nu}+\bar{\Theta}_{A}\,^{*}\! F^{(1)\mu\nu}a_{\nu}\Big]\Big\}\:.\nonumber 
\end{eqnarray}
Here all co-variant derivatives $\bar{\nabla}$ and raising and lowering
indices are with respect to the background metric $\bar{g}_{\mu\nu}$,
with $h\equiv h_{\mu}^{\mu}$ and $F_{\mu\nu}^{(1)}\equiv\bar{\nabla}_{[\mu}a_{\nu]}$.%
\footnote{In this paper we define the symmetrization $A_{(\mu}B_{\nu)}\equiv A_{\mu}B_{\nu}+A_{\nu}B_{\mu}$
and the anti-symmetrization $A_{[\mu}B_{\nu]}\equiv A_{\mu}B_{\nu}-A_{\nu}B_{\mu}$
without the factor of $\frac{1}{2}$.%
} These actions are written as integrals of total derivatives, which
means they are boundary terms.

\subsection{Equations of Motion and Background}

The EOMs are 
\begin{eqnarray}
R_{\mu\nu}-\frac{1}{2}Rg_{\mu\nu}+\Lambda g_{\mu\nu}+\lambda C_{\mu\nu} & = & \frac{1}{2}T_{\mu\nu}\:,\\
\nabla_{\mu}\left(F^{\mu\nu}-\lambda_{A}\bar{\Theta}_{A}\,^{*}\!\bar{F}^{\mu\nu}\right) & = & 0\:,\\
\nabla^{2}\vartheta-\frac{\delta V}{\delta\vartheta}-\frac{\lambda}{4}\frac{\delta\Theta}{\delta\vartheta}\,^{*}\! RR+\frac{\lambda_{A}}{4}\frac{\delta\Theta_{A}}{\delta\vartheta}\,^{*}\! FF & = & 0\:,
\end{eqnarray}
where 
\begin{eqnarray}
C_{\mu\nu} & = & \frac{1}{2}\nabla^{\alpha}\nabla^{\beta}\left(\Theta\,^{*}\! R_{\alpha\mu\beta\nu}+\Theta\,^{*}\! R_{\alpha\nu\beta\mu}\right)\nonumber \\
 & = & \frac{1}{2}\left[\left(\nabla^{\alpha}\nabla^{\beta}\Theta\right)\,^{*}\! R_{\alpha\mu\beta\nu}+\left(\nabla^{\beta}\Theta\right)\epsilon_{\beta\mu\gamma\delta}\left(\nabla^{\gamma}R_{\nu}^{\delta}\right)\right]+\left(\mu\leftrightarrow\nu\right)
\end{eqnarray}
and 
\begin{eqnarray}
T_{\mu\nu} & = & \left[F_{\mu\rho}F_{\nu}^{\phantom{\nu}\rho}-\frac{1}{4}g_{\mu\nu}F^{2}\right]+\left[\left(\nabla_{\mu}\vartheta\right)\left(\nabla_{\nu}\vartheta\right)-\frac{1}{2}g_{\mu\nu}\left(\nabla\vartheta\right)^{2}-g_{\mu\nu}V[\vartheta]\right]\:.
\end{eqnarray}

We choose the background ansatz to be 
\begin{equation}
\left\{ \begin{aligned} & d\bar{s}^{2}=-F(z)dt^{2}+\frac{dz^{2}}{F(z)}+r(z)^{2}\left(dx^{2}+dy^{2}\right)\\
 & \bar{A}=\Phi(z)dt\:,\qquad\bar{\theta}=\theta(z)
\end{aligned}
\:.\right.\label{Background}
\end{equation}
The temperature is given by $F(z)=4\pi T(z-z_{H})+O\left((z-z_{H})^{2}\right)$
and $\Phi(z)=O(z-z_{H})$ near the horizon $z=z_{H}$. The boundary
is at $z=\infty$.

The background EOMs are 
\begin{eqnarray}
\frac{d^{2}r(z)}{dz^{2}}+\frac{1}{4}r(z)\left(\frac{d\theta(z)}{dz}\right)^{2} & = & 0\:,\label{eq:BackgroundEOM_r(z)}\\
\frac{d^{2}F(z)}{dz^{2}}-\frac{2F(z)}{r(z)^{2}}\left(\frac{dr(z)}{dz}\right)^{2}-\left(\frac{d\Phi(z)}{dz}\right)^{2}+\frac{F(z)}{2}\left(\frac{d\theta(z)}{dz}\right)^{2} & = & 0\:,\label{eq:BackgroundEOM_F(z)}\\
\frac{d^{2}\Phi(z)}{dz^{2}}+\frac{2}{r(z)}\frac{dr(z)}{dz}\frac{d\Phi(z)}{dz} & = & 0\:,\label{eq:BackgroundEOM_Phi(z)}\\
\frac{d}{dz}\left[r(z)^{2}F(z)\left(\frac{d}{dz}\theta(z)\right)\right]-r(z)^{2}\frac{\delta V\left[\theta\right]}{\delta\theta} & = & 0\:.\label{AxionEOM}
\end{eqnarray}
and a constraint equation from the trace of Einstein equation: 
\begin{align}
\frac{d^{2}F(z)}{dz^{2}}+4\frac{F(z)}{r(z)}\frac{d^{2}r(z)}{dz^{2}}+2\frac{F(z)}{r(z)^{2}}\left(\frac{dr(z)}{dz}\right)^{2}+\frac{4}{r(z)}\frac{dF(z)}{dz}\frac{dr(z)}{dz}\quad\nonumber \\
+\frac{1}{2}F(z)\left(\frac{d\theta(z)}{dz}\right)^{2}+2V\left[\theta\right]=\frac{12}{L^{2}}\:.
\end{align}

\bigskip{}

\section{Vector Mode Fluctuations and Angular Momentum Density}

\subsection{Formula for Angular Momentum Density}

In this section, we follow the method proposed in \cite{Paper1} to
calculate the angular momentum density. The gauge conditions are chosen
to be $h_{\mu z}=a_{z}=0$. We need to study only the static case,
so all fluctuations are time-independent. For completeness, we first
review the derivation of the formula used to compute angular momentum
density. The metric at the boundary is $\gamma_{\alpha\beta}=\eta_{\alpha\beta}+\delta\gamma_{\alpha\beta}$,
where $\eta_{\alpha\beta}=(-1,1,1)$ is the flat Lorentzian metric
and $\delta\gamma_{\alpha\beta}=\bar{h}_{\alpha\beta}$ is the metric
fluctuation at the boundary. $\alpha,\beta=t,x,y$ and $i,j,k=x,y$.
The energy-stress tensor is defined as 
\begin{equation}
T^{\alpha\beta}=\frac{2}{\sqrt{-\gamma}}\frac{\delta S}{\delta\gamma_{\alpha\beta}}\Bigg|_{\gamma_{\alpha\beta}=\eta_{\alpha\beta}}\:.
\end{equation}
Then the linearized (first order) on-shell action we calculate from
holography will in general takes the form 
\begin{equation}
S^{(1)}=\frac{1}{2}\int d^{3}x\bar{h}_{\alpha\beta}(x)T^{\alpha\beta}(x)\:.\label{eq:S1_hT}
\end{equation}
The $\beta=t$ component of the conservation law $\nabla_{\alpha}T^{\alpha\beta}=0$
in the static case and flat background reads 
\begin{equation}
\partial_{i}T^{ti}(\vec{x})=0\:,
\end{equation}
which has a general solution 
\begin{equation}
T^{ti}(\vec{x})=\frac{1}{2}\epsilon^{ij}\partial_{j}\ell(\vec{x})\:,\label{eq:Tti_l}
\end{equation}
where $\epsilon^{xy}=-\epsilon^{yx}=1$, $\epsilon^{xx}=\epsilon^{yy}=0$
and $\ell(\vec{x})$ is an arbitrary function. It is straightforward
to see that $\ell(\vec{x})$ is the angular momentum density by definition:
\begin{equation}
L=\int d^{2}\vec{x}\epsilon_{ij}x^{i}T^{tj}(\vec{x})=\frac{1}{2}\int d^{2}\vec{x}\epsilon_{ij}x^{i}\epsilon^{jk}\partial_{k}\ell(\vec{x})=\int_{V}d^{2}\vec{x}\ell(\vec{x})\:,
\end{equation}
where in the last step we have integrated by parts, assumed that the
system and hence $\ell(\vec{x})$ are localized in a finite volume
$V$, and used $\epsilon_{ij}\epsilon^{jk}=-\delta_{i}^{k}$. At the
end the volume $V$ can be extended to include the whole space. Plug
(\ref{eq:Tti_l}) into (\ref{eq:S1_hT}), turn on only $\bar{h}_{ti}(\vec{x})$
fluctuation and integrate by parts, we get 
\begin{equation}
S^{(1)}=\frac{1}{2}\int d^{3}x\ell(\vec{x})\epsilon^{ij}\partial_{i}\bar{h}_{tj}(\vec{x})\:.\label{eq:AngMomFormula_1}
\end{equation}
When the system is homogeneous, $\ell(\vec{x})=\ell$ is a constant
and can be factored out of the integral. Then we have 
\begin{equation}
S^{(1)}=\frac{\ell}{2}\int d^{3}x\left(\frac{\partial}{\partial x}\bar{h}_{ty}(\vec{x})-\frac{\partial}{\partial y}\bar{h}_{tx}(\vec{x})\right)\:.\label{eq:AngMomFormula_2}
\end{equation}
(\ref{eq:AngMomFormula_1}) and (\ref{eq:AngMomFormula_2}) are the
template formulae for computing angular momentum density in holography.

\subsection{First Order On-Shell Action}

We now calculate the linearized on-shell action $S^{(1)}=S_{\textrm{bulk}}^{(1)}+S_{\textrm{GH}}^{(1)}+S_{\partial\textrm{CS}}^{(1)}+S_{\textrm{ct}}^{(1)}$.
The first part of the contribution is from the z-derivative term in
$S_{\textrm{bulk}}^{(1)}$ plus the boundary terms: 
\begin{align}
 & \frac{1}{2\kappa^{2}}\int_{z=\infty}d^{3}x\Bigg\{2r(z)\left(F(z)\frac{dr(z)}{dz}-\frac{r(z)}{L}\sqrt{F(z)}\right)h_{t}^{t}(\vec{x},z)+r(z)^{2}\frac{d\Phi(z)}{dz}a_{t}(\vec{x},z)\nonumber \\
 & \qquad+r(z)\left(F(z)\frac{dr(z)}{dz}+\frac{r(z)}{2}\frac{dF(z)}{dz}-\frac{2r(z)}{L}\sqrt{F(z)}\right)\left(h_{x}^{x}(\vec{x},z)+h_{y}^{y}(\vec{x},z)\right)\nonumber \\
 & \qquad-r(z)^{2}F(z)\frac{d\theta(z)}{dz}\delta\theta(\vec{x},z)\label{S1_part1_a}\\
 & \qquad+\frac{\lambda}{2}\bar{\Theta}(z)\frac{dr(z)}{dz}\left(2F(z)\frac{dr(z)}{dz}-r(z)\frac{dF(z)}{dz}\right)\left(\frac{\partial}{\partial x}h_{t}^{y}(\vec{x},z)-\frac{\partial}{\partial y}h_{t}^{x}(\vec{x},z)\right)\nonumber \\
 & \qquad+\frac{\lambda}{4}\bar{\Theta}(z)r(z)\left(2F(z)\frac{dr(z)}{dz}-r(z)\frac{dF(z)}{dz}\right)\left(\frac{\partial^{2}}{\partial z\partial x}h_{t}^{y}(\vec{x},z)-\frac{\partial^{2}}{\partial z\partial y}h_{t}^{x}(\vec{x},z)\right)\Bigg\}\:.\nonumber 
\end{align}
Using the asymptotic behavior of the metric: 
\begin{equation}
\left\{ \begin{aligned} & r(z)=\frac{z}{L}+O\left(\frac{1}{z^{3}}\right)\\
 & F(z)=\left(\frac{z}{L}\right)^{2}+\frac{\Gamma}{z}+O\left(\frac{1}{z^{2}}\right)\\
 & \Phi(z)=\Phi_{0}+\frac{\Phi_{1}}{z}+O\left(\frac{1}{z^{2}}\right)
\end{aligned}
\:,\right.\label{Metric_NB}
\end{equation}
the first two lines of the integrand in (\ref{S1_part1_a}) are 
\[
\frac{\Gamma}{2L^{2}}\left(2\bar{h}_{t}^{t}-\bar{h}_{x}^{x}-\bar{h}_{y}^{y}\right)-\frac{\Phi_{1}}{L^{2}}\bar{a}_{t}\:.
\]
Since we do not turn on these boundary fields, they have no contribution.
The axion has near-boundary behavior 
\begin{equation}
\theta(z)=\theta_{0}z^{-\Delta_{-}}\ldots+\theta_{1}z^{-\Delta_{+}}\ldots\:,\label{Axion_Asymptotic}
\end{equation}
where 
\begin{equation}
\Delta_{\pm}=\frac{3}{2}\pm\sqrt{\frac{9}{4}+m^{2}L^{2}}\:.\label{ConformalDimension}
\end{equation}
The first coefficient $\theta_{0}$ is equal to the source $J$ and
the second coefficient $\theta_{1}$ equal to the condensate $\langle\mathcal{O}\rangle$.
Since we are considering a sourceless case, we set $\theta_{0}=0$.
The axion fluctuation $\delta\theta$ has similar near-boundary behavior:
\[
\delta\theta=\delta\bar{\theta}z^{-\Delta_{-}}+\ldots\:,
\]
then the third line in (\ref{S1_part1_a}) becomes 
\[
\frac{\Delta_{+}}{L^{4}}\theta_{1}\delta\bar{\theta}\:,
\]
which has no contribution because $\delta\bar{\theta}$ is turned
off. Thus the first part's contribution is only from the last two
lines in (\ref{S1_part1_a}): 
\begin{align}
 & \frac{\lambda}{4\kappa^{2}}\int_{z=\infty}d^{3}x\Bigg\{\bar{\Theta}(z)\frac{dr(z)}{dz}\left(2F(z)\frac{dr(z)}{dz}-r(z)\frac{dF(z)}{dz}\right)\left(\frac{\partial}{\partial x}h_{t}^{y}(\vec{x},z)-\frac{\partial}{\partial y}h_{t}^{x}(\vec{x},z)\right)\nonumber \\
 & \qquad+\frac{1}{2}\bar{\Theta}(z)r(z)\left(2F(z)\frac{dr(z)}{dz}-r(z)\frac{dF(z)}{dz}\right)\left(\frac{\partial^{2}}{\partial z\partial x}h_{t}^{y}(\vec{x},z)-\frac{\partial^{2}}{\partial z\partial y}h_{t}^{x}(\vec{x},z)\right)\Bigg\}\:.\label{S1_part1_b}
\end{align}
The second part of the contribution to $S^{(1)}$ is from the $x$-
and $y$-derivative terms in $S_{\textrm{bulk}}^{(1)}$. The part
involving scalar and tensor mode fluctuations are quadratic in spatial
derivatives: 
\[
\frac{1}{2\kappa^{2}}\int d^{4}x\Bigg\{-\left(\frac{\partial^{2}}{\partial x^{2}}+\frac{\partial^{2}}{\partial y^{2}}\right)h_{t}^{t}(\vec{x},z)-\frac{\partial^{2}}{\partial y^{2}}h_{x}^{x}(\vec{x},z)-\frac{\partial^{2}}{\partial x^{2}}h_{y}^{y}(\vec{x},z)+2\frac{\partial^{2}}{\partial x\partial y}h_{y}^{x}(\vec{x},z)\Bigg\}\:,
\]
thus is of higher order. The part involving vector mode fluctuations
will give the main contribution: 
\begin{align*}
 & \frac{\lambda}{4\kappa^{2}}\int d^{4}x\Bigg\{\Bigg[\left(r(z)^{2}\frac{d}{dz}\left(\frac{1}{r(z)}\frac{dr(z)}{dz}\frac{dF(z)}{dz}\right)-4F(z)\frac{dr(z)}{dz}\frac{d^{2}r(z)}{dz^{2}}\right)\bar{\Theta}(z)\\
 & \qquad+r(z)^{3}\left(\frac{d}{dz}\frac{F(z)}{r(z)^{2}}\right)\frac{dr(z)}{dz}\frac{d\bar{\Theta}(z)}{dz}\Bigg]\left(\frac{\partial}{\partial x}h_{t}^{y}(\vec{x},z)-\frac{\partial}{\partial y}h_{t}^{x}(\vec{x},z)\right)\\
 & \qquad+r(z)\bar{\Theta}(z)\left[r(z)^{2}\frac{d}{dz}\left(\frac{1}{r(z)}\frac{dF(z)}{dz}\right)-2F(z)\frac{d^{2}r(z)}{dz^{2}}\right]\left(\frac{\partial^{2}}{\partial z\partial x}h_{t}^{y}(\vec{x},z)-\frac{\partial^{2}}{\partial z\partial y}h_{t}^{x}(\vec{x},z)\right)\Bigg\}\\
 & +\frac{\lambda_{A}}{2\kappa^{2}}\int d^{4}x\Bigg\{\bar{\Theta}_{A}(z)\frac{d\Phi(z)}{dz}\left(\frac{\partial}{\partial x}a_{y}(\vec{x},z)-\frac{\partial}{\partial y}a_{x}(\vec{x},z)\right)\Bigg\}\:.
\end{align*}
The quantity in the first $\left[\ldots\right]$ is a total derivative,
so this part can be integrated by parts, which gives
\begin{align*}
 & \frac{\lambda}{4\kappa^{2}}\int d^{3}x\bar{\Theta}(z)\frac{dr(z)}{dz}\left(r(z)\frac{dF(z)}{dz}-2F(z)\frac{dr(z)}{dz}\right)\left(\frac{\partial}{\partial x}h_{t}^{y}(\vec{x},z)-\frac{\partial}{\partial y}h_{t}^{x}(\vec{x},z)\right)\Bigg|_{z=z_{H}}^{z=\infty}\\
 & +\frac{\lambda}{4\kappa^{2}}\int d^{4}x\Bigg\{\bar{\Theta}(z)\Bigg[-\frac{dr(z)}{dz}\left(r(z)\frac{dF(z)}{dz}-2F(z)\frac{dr(z)}{dz}\right)+r(z)^{3}\frac{d}{dz}\left(\frac{1}{r(z)}\frac{dF(z)}{dz}\right)\\
 & \qquad-2r(z)F(z)\frac{d^{2}r(z)}{dz^{2}}\Bigg]\left(\frac{\partial^{2}}{\partial z\partial x}h_{t}^{y}(\vec{x},z)-\frac{\partial^{2}}{\partial z\partial y}h_{t}^{x}(\vec{x},z)\right)\Bigg\}\\
 & +\frac{\lambda_{A}}{2\kappa^{2}}\int d^{4}x\Bigg\{\bar{\Theta}_{A}(z)\frac{d\Phi(z)}{dz}\left(\frac{\partial}{\partial x}a_{y}(\vec{x},z)-\frac{\partial}{\partial y}a_{x}(\vec{x},z)\right)\Bigg\}\:.
\end{align*}
Combine this with (\ref{S1_part1_b}), we have:
\begin{align}
S^{(1)}= & \frac{\lambda}{4\kappa^{2}}\int_{z=z_{H}}d^{3}x\bar{\Theta}(z)\frac{dr(z)}{dz}\left(2F(z)\frac{dr(z)}{dz}-r(z)\frac{dF(z)}{dz}\right)\left(\frac{\partial}{\partial x}h_{t}^{y}(\vec{x},z)-\frac{\partial}{\partial y}h_{t}^{x}(\vec{x},z)\right)\nonumber \\
+ & \frac{\lambda}{8\kappa^{2}}\int_{z=\infty}d^{3}x\Bigg\{\bar{\Theta}(z)r(z)\left(2F(z)\frac{dr(z)}{dz}-r(z)\frac{dF(z)}{dz}\right)\left(\frac{\partial^{2}}{\partial z\partial x}h_{t}^{y}(\vec{x},z)-\frac{\partial^{2}}{\partial z\partial y}h_{t}^{x}(\vec{x},z)\right)\Bigg\}\nonumber \\
+ & \frac{\lambda}{4\kappa^{2}}\int d^{4}x\Bigg\{\bar{\Theta}(z)\Bigg[-\frac{dr(z)}{dz}\left(r(z)\frac{dF(z)}{dz}-2F(z)\frac{dr(z)}{dz}\right)+r(z)^{3}\frac{d}{dz}\left(\frac{1}{r(z)}\frac{dF(z)}{dz}\right)\nonumber \\
 & \qquad-2r(z)F(z)\frac{d^{2}r(z)}{dz^{2}}\Bigg]\left(\frac{\partial^{2}}{\partial z\partial x}h_{t}^{y}(\vec{x},z)-\frac{\partial^{2}}{\partial z\partial y}h_{t}^{x}(\vec{x},z)\right)\Bigg\}\label{S1_tot}\\
+ & \frac{\lambda_{A}}{2\kappa^{2}}\int d^{4}x\Bigg\{\bar{\Theta}_{A}(z)\frac{d\Phi(z)}{dz}\left(\frac{\partial}{\partial x}a_{y}(\vec{x},z)-\frac{\partial}{\partial y}a_{x}(\vec{x},z)\right)\Bigg\}\:.\nonumber 
\end{align}
This is the linearized action we will use in the next subsection to
compute angular momentum.

\subsection{Vector Mode Fluctuations and Angular Momentum Density}

Since $S^{(1)}$ is already linear in spatial derivatives, we only
need to solve equations for $h_{t}^{i}(\vec{x},z)$ ($i=x,y$) at
the homogeneous leading order. That is, we can view spatial derivatives
as small quantities and solve only up to leading order in derivative
expansion. There are four relevant equations for vector mode fluctuations:
the $tz$-component of the linearized Einstein equation and the $z$-component
of the linearized Maxwell equation 
\begin{eqnarray}
\frac{F(z)}{2r(z)^{2}}\frac{\partial}{\partial z}\left(\frac{r(z)^{2}}{F(z)}\partial_{i}h_{t}^{i}(\vec{x},z)\right) & = & O\left(\vec{\partial}^{2}\right)\:,\label{eq:LinearEinsteinEq_tz}\\
\frac{d\Phi(z)}{dz}\partial_{i}h_{t}^{i}(\vec{x},z)+\frac{F(z)}{r(z)^{2}}\frac{\partial}{\partial z}\left(\partial_{i}a_{i}(\vec{x},z)\right) & = & 0\:,\label{eq:LinearMaxwellEq_z}
\end{eqnarray}
where sum over $i=x,y$ is understood, and the $ti$-component of
the linearized Einstein equation and $i$-component of the linearized
Maxwell equation
\begin{eqnarray}
\frac{d}{dz}\left[r(z)^{4}\left(\frac{d}{dz}h_{t}^{i}(\vec{x},z)\right)+r(z)^{2}\frac{d\Phi(z)}{dz}a_{i}(\vec{x},z)\right] & = & O\left(\vec{\partial}\right)\:,\label{eq:LinearEinsteinEq_ti}\\
r(z)^{2}\frac{d\Phi(z)}{dz}\left(\frac{d}{dz}h_{t}^{i}(\vec{x},z)\right)+\frac{d}{dz}\left[F(z)\frac{d}{dz}a_{i}(\vec{x},z)\right] & = & O\left(\vec{\partial}\right)\:.\label{eq:LinearMaxwellEq_i}
\end{eqnarray}
Equation (\ref{eq:LinearEinsteinEq_tz}) can be directly integrated
out, which gives 
\[
\frac{r(z)^{2}}{F(z)}\partial_{i}h_{t}^{i}(\vec{x},z)=C_{1}(\vec{x})+O\left(\vec{\partial}^{2}\right)\:,
\]
where $C_{1}(\vec{x})$ is an arbitrary function independent of $z$.
Since the right hand side is already independent of $z$, the left
hand side must also be independent of $z$. To achieve this, the $z$-dependence
of $h_{t}^{i}(\vec{x},z)$ must cancel the prefactor $r(z)^{2}/F(z)$.
Noticing that this factor goes to $1$ at the boundary, and we want
to normalize $h_{t}^{i}(\vec{x},z)$ at the boundary as $h_{t}^{i}(\vec{x},z\rightarrow\infty)=\bar{h}_{ti}(\vec{x})$,
we get the solution 
\begin{equation}
h_{t}^{i}(\vec{x},z)=\frac{F(z)}{r(z)^{2}}\bar{h}_{ti}(\vec{x})+O\left(\vec{\partial}\right)\:.\label{eq:VectorModeSolutions_hti}
\end{equation}
Here indices of boundary fields $\bar{h}_{ti}$ are raised and lowered
by the 3-d flat Lorentzian metric $\eta_{\alpha\beta}=\left(-1,1,1\right)$.
$C_{1}(\vec{x})$ is then determined accordingly. Plug this solution
into equation (\ref{eq:LinearMaxwellEq_z}) and integrate in out,
we get 
\[
\partial_{i}\left(a_{i}(\vec{x},z)+\Phi(z)\bar{h}_{ti}(\vec{x})\right)=C_{2}(\vec{x})+O\left(\vec{\partial}^{2}\right)\:,
\]
where $C_{2}(\vec{x})$ is an arbitrary function independent of $z$.
Same as before, we want the left hand side to be $z$-independent.
Noticing that $\Phi(z\rightarrow\infty)=\Phi_{0}+O\left(z^{-1}\right)$
and we want to normalize $a_{i}(\vec{x},z)$ at the boundary as $a_{i}(\vec{x},z\rightarrow\infty)=0$,
we get the solution 
\begin{equation}
a_{i}(\vec{x},z)=\left(\Phi_{0}-\Phi(z)\right)\bar{h}_{ti}(\vec{x})+O\left(\vec{\partial}\right)\:.\label{eq:VectorModeSolutions_ai}
\end{equation}
Using background EOMs it is straightforward to check that the solutions
(\ref{eq:VectorModeSolutions_hti}) and (\ref{eq:VectorModeSolutions_ai})
solve the second order equations (\ref{eq:LinearEinsteinEq_ti}) and
(\ref{eq:LinearMaxwellEq_i}) as well. Thus the vector mode fluctuations
are completely solved at the leading order in derivative expansion.

Plug these solutions into (\ref{S1_tot}), and integrate by parts
the bulk integrals, we arrive at 
\begin{eqnarray*}
S^{(1)} & = & -\frac{\lambda_{A}}{4\kappa^{2}}\int_{z=\infty}d^{3}x\bar{\Theta}_{A}(z)\left(\Phi_{0}-\Phi(z)\right)^{2}\left(\frac{\partial}{\partial x}\bar{h}_{ty}(\vec{x})-\frac{\partial}{\partial y}\bar{h}_{tx}(\vec{x})\right)\\
 &  & +\frac{\lambda}{8\kappa^{2}}\int_{z=z_{H}}d^{3}x\bar{\Theta}(z)\frac{dF(z)}{dz}\left(2\frac{F(z)}{r(z)}\frac{dr(z)}{dz}-\frac{dF(z)}{dz}\right)\left(\frac{\partial}{\partial x}\bar{h}_{ty}(\vec{x})-\frac{\partial}{\partial y}\bar{h}_{tx}(\vec{x})\right)\\
 &  & +\frac{\lambda_{A}}{4\kappa^{2}}\int_{z=z_{H}}d^{3}x\bar{\Theta}_{A}(z)\left(\Phi_{0}-\Phi(z)\right)^{2}\left(\frac{\partial}{\partial x}\bar{h}_{ty}(\vec{x})-\frac{\partial}{\partial y}\bar{h}_{tx}(\vec{x})\right)\\
 &  & -\frac{\lambda}{8\kappa^{2}}\int d^{4}xr(z)^{4}\left[\frac{d}{dz}\left(\frac{F(z)}{r(z)^{2}}\right)\right]^{2}\frac{d\bar{\Theta}(z)}{dz}\left(\frac{\partial}{\partial x}\bar{h}_{ty}(\vec{x})-\frac{\partial}{\partial y}\bar{h}_{tx}(\vec{x})\right)\\
 &  & +\frac{\lambda_{A}}{4\kappa^{2}}\int d^{4}x\left(\Phi_{0}-\Phi(z)\right)^{2}\frac{d\bar{\Theta}_{A}(z)}{dz}\left(\frac{\partial}{\partial x}\bar{h}_{ty}(\vec{x})-\frac{\partial}{\partial y}\bar{h}_{tx}(\vec{x})\right)\:.
\end{eqnarray*}
Noticing $F(z)=4\pi T(z-z_{H})+O\left((z-z_{H})^{2}\right)$ and $\Phi(z)=O\left(z-z_{H}\right)$
near the horizon and $\Phi(z\rightarrow\infty)=\Phi_{0}+O\left(z^{-1}\right)$
near the boundary, we get 
\begin{eqnarray*}
S^{(1)} & = & \frac{1}{2\kappa^{2}}\left(\frac{1}{2}\Phi_{0}^{2}\lambda_{A}\bar{\Theta}_{A}(z_{H})-4\pi^{2}T^{2}\lambda\bar{\Theta}(z_{H})\right)\int d^{3}x\left(\frac{\partial}{\partial x}\bar{h}_{ty}(\vec{x})-\frac{\partial}{\partial y}\bar{h}_{tx}(\vec{x})\right)\\
 &  & +\frac{1}{2\kappa^{2}}\int_{z_{H}}^{\infty}dz\left[\frac{\lambda_{A}}{2}\left(\Phi_{0}-\Phi(z)\right)^{2}\frac{d\bar{\Theta}_{A}(z)}{dz}\right]\int d^{3}x\left(\frac{\partial}{\partial x}\bar{h}_{ty}(\vec{x})-\frac{\partial}{\partial y}\bar{h}_{tx}(\vec{x})\right)\\
 &  & +\frac{1}{2\kappa^{2}}\int_{z_{H}}^{\infty}dz\left\{ -\frac{\lambda}{4}r(z)^{4}\left[\frac{d}{dz}\left(\frac{F(z)}{r(z)^{2}}\right)\right]^{2}\frac{d\bar{\Theta}(z)}{dz}\right\} \int d^{3}x\left(\frac{\partial}{\partial x}\bar{h}_{ty}(\vec{x})-\frac{\partial}{\partial y}\bar{h}_{tx}(\vec{x})\right)\:.
\end{eqnarray*}
Compare with the formula (\ref{eq:AngMomFormula_2}), we get 
\begin{eqnarray}
\ell & = & \frac{\lambda_{A}}{2\kappa^{2}}\left\{ \Phi_{0}^{2}\Theta_{A}\left[\theta(z_{H})\right]+\int_{z_{H}}^{\infty}dz\left(\Phi_{0}-\Phi(z)\right)^{2}\frac{d\Theta_{A}\left[\theta(z)\right]}{dz}\right\} \nonumber \\
 &  & -\frac{\lambda}{4\kappa^{2}}\left\{ 16\pi^{2}T^{2}\Theta\left[\theta(z_{H})\right]+\int_{z_{H}}^{\infty}dz\, r(z)^{4}\left[\frac{d}{dz}\left(\frac{F(z)}{r(z)^{2}}\right)\right]^{2}\frac{d\Theta\left[\theta(z)\right]}{dz}\right\} \:,\label{AngMomDensity}
\end{eqnarray}
where $\Phi_{0}=\Phi(z=\infty)$.

This result is in agreement with that obtained in \cite{Liu:2013cha}.%
\footnote{To go from our expression to that of \cite{Liu:2013cha}, one first
need to substitute in the following field and coupling redefinition:
\[
A_{\mu}\Rightarrow LA_{\mu}\:,\qquad\lambda_{A}\Rightarrow-4\beta_{\mathrm{CS}}\:,\qquad\lambda\Rightarrow\alpha_{\mathrm{CS}}L^{2}\:,\qquad\Theta_{A}\left[\vartheta\right],\Theta\left[\vartheta\right]\Rightarrow\vartheta\:,
\]
where on the right hand side of ``$\Rightarrow$'' are the notations
of \cite{Liu:2013cha} (the AdS radius $L$ is denoted by $\ell$
there, but to avoid confusion, we will still use $L$ here). This
implies $\Phi_{0}\Rightarrow L\mu$. After these substitution, our
actions take exactly the same form as those used in \cite{Liu:2013cha}.
Next, to transform the metric to that of \cite{Liu:2013cha}, we redefine
the coordinate system as 
\[
r(z)=\frac{L}{\xi}\:,\qquad F(z)=f(\xi)\frac{L^{2}}{\xi^{2}}\:,\qquad z=-L^{2}\int\frac{\sqrt{f(\xi)h(\xi)}}{\xi^{2}}d\xi\:.
\]
Now one can check our metric (\ref{Background}) takes the form of
eq. (1.4) in \cite{Liu:2013cha}, where their AdS radius coordinate
``$z$'' is re-denoted by $\xi$ here to avoid conflict of symbols.
Next it is straightforward to substitute these expressions into (\ref{AngMomDensity})
and see that it takes exactly the same form as eqs. (1.5) and (1.6)
in \cite{Liu:2013cha}. Notice that the upper integral limit of $\infty$
here shall be replaced by $0$ as $\xi=0$ is the AdS boundary in
\cite{Liu:2013cha}.%
} The basic concept behind the method here and that in \cite{Liu:2013cha}
are the same, which is to look at the momentum density 1-point function's
response to spatially inhomogeneous perturbations. But the executions
of the computation are done in different ways. In \cite{Liu:2013cha}
the inhomogeneous momentum density are obtained by directly solving
inhomogeneous bulk equations of motion. While here by looking at first
order action's response rather than that of momentum density itself
and performing an integration by parts, we obtain the template formula
(\ref{eq:AngMomFormula_2}). The spatial derivative is shifted from
the momentum density to the metric fluctuation. Then to calculate
the homogeneous angular momentum density, technically we only need
to solve homogeneous bulk equations which is very easy to do. The
results of course agree as we have checked explicitly here, because
the integration by parts in the action is just a mathematical trick
and should not have any physical consequence.

\subsection{Effective Masses and Membrane Paradigm Violation}

From (\ref{AngMomDensity}), we can see that in general both gauge
and gravitational Chern-Simons terms contribute to the angular momentum
density. The contributions are not just from the horizon area, as
opposed to many transport coefficients such as the shear and Hall
viscosities, which we will compute in the next section. The fact that
parts of the contribution are written as integrals from the horizon
to the boundary suggests that the IR degrees of freedom interact non-trivially
with the UV degrees of freedom to generate the angular momentum density.
In the so-called membrane paradigm \cite{Iqbal:2008by}, many zero-frequency
and zero-momentum linear response transport coefficients can be expressed
completely in terms of geometric quantities evaluated at the black
hole horizon in holographic dual. But here the angular momentum density
(\ref{AngMomDensity}) is clearly an exception. One key ingredient
in the derivation of the membrane paradigm in \cite{Iqbal:2008by}
is that the bulk degrees of freedom associated with the linear response
in question is massless, thus the bulk equations of motion can be
integrated out which yields the membrane paradigm. But when a mass
term is included (or generated) in the equations of motion, it usually
spoils the integrability thus can break the membrane paradigm. This
is what happens here to the angular momentum density. Although the
vector mode fluctuations we are considering here originate from massless
bulk fluctuations, they acquire effective $z$-dependent masses spontaneously
from the non-trivial profile of $\Phi(z)$ and the geometry. We will
show in the following that equations (\ref{eq:LinearEinsteinEq_ti})
and (\ref{eq:LinearMaxwellEq_i}) are actually equations of motion
for massive vector fluctuations.

First, we notice that the background equation (\ref{eq:BackgroundEOM_Phi(z)})
can be solved formally: 
\begin{equation}
\Phi(z)=\Phi_{0}-\frac{\Phi_{1}}{L^{2}}\int_{\infty}^{z}\frac{d\xi}{r(\xi)^{2}}\:,\qquad\textrm{with}\qquad\Phi_{1}=-\Phi_{0}L^{2}\left(\int_{z_{H}}^{\infty}\frac{d\xi}{r(\xi)^{2}}\right)^{-1}\:.
\end{equation}
In fact, up to normalization factors, $\Phi_{0}=\mu$ and $\Phi_{1}=-\rho$
where $\mu$ and $\rho$ are the chemical potential and charge density
of the system. Now using the properties of the solutions (\ref{eq:VectorModeSolutions_hti})
and (\ref{eq:VectorModeSolutions_ai}) 
\begin{eqnarray*}
a_{i}(\vec{x},z) & = & \frac{r(z)^{2}}{F(z)}\left[\Phi_{0}-\Phi(z)\right]h_{t}^{i}(\vec{x},z)+O\left(\vec{\partial}\right)\:,\\
\frac{\partial}{\partial z}h_{t}^{i}(\vec{x},z) & = & h_{t}^{i}(\vec{x},z)\frac{\partial}{\partial z}\log\frac{F(z)}{r(z)^{2}}+O\left(\vec{\partial}\right)\:,
\end{eqnarray*}
we can manipulate the second term in (\ref{eq:LinearEinsteinEq_ti})
to get the following form 
\begin{equation}
\frac{d}{dz}\left[r(z)^{4}\left(\frac{d}{dz}h_{t}^{i}(\vec{x},z)\right)\right]-\frac{\Phi_{1}^{2}L^{-4}}{F(z)}h_{t}^{i}(\vec{x},z)=O\left(\vec{\partial}\right)\:.
\end{equation}
In this equation, the first term is the equation of motion for massless
vector mode fluctuation of the metric as usual, but the second term
corresponds to an effective mass term%
\footnote{To read off the value of mass, this can be compared, for example,
with eqs. (2.49) and (2.50) in \cite{deRham:2014zqa}.%
} with a $z$-dependent mass square: 
\begin{equation}
m_{h}^{2}(z)=\frac{\Phi_{1}^{2}L^{-4}}{r(z)^{4}}\:.
\end{equation}
Similarly manipulating the first term in (\ref{eq:LinearMaxwellEq_i})
it becomes 
\begin{equation}
\frac{d}{dz}\left[F(z)\frac{d}{dz}a_{i}(\vec{x},z)\right]-\left[\frac{d}{dz}\left(\frac{F(z)}{r(z)^{2}}\right)\right]\left(\int_{\infty}^{z}\frac{d\xi}{r(\xi)^{2}}\right)^{-1}a_{i}(\vec{x},z)=O\left(\vec{\partial}\right)\:.
\end{equation}
The first term in the above equation corresponds to that of massless
Maxwell field and the second term an effective Proca mass term with
a $z$-dependent mass square: 
\begin{equation}
m_{a}^{2}(z)=\left[\frac{d}{dz}\left(\frac{F(z)}{r(z)^{2}}\right)\right]\left(\int_{\infty}^{z}\frac{d\xi}{r(\xi)^{2}}\right)^{-1}\:.
\end{equation}
Then (\ref{AngMomDensity}) can be written as 
\begin{align}
\ell= & \frac{\lambda_{A}}{2\kappa^{2}}\left\{ \Phi_{0}^{2}\Theta_{A}\left[\theta(z_{H})\right]+\int_{z_{H}}^{\infty}dz\, m_{h}^{2}(z)\, r(z)^{4}\left(\int_{\infty}^{z}\frac{d\xi}{r(\xi)^{2}}\right)^{2}\frac{d\Theta_{A}\left[\theta(z)\right]}{dz}\right\} \nonumber \\
 & -\frac{\lambda}{4\kappa^{2}}\left\{ 16\pi^{2}T^{2}\Theta\left[\theta(z_{H})\right]+\int_{z_{H}}^{\infty}dz\left[m_{a}^{2}(z)\right]^{2}r(z)^{4}\left(\int_{\infty}^{z}\frac{d\xi}{r(\xi)^{2}}\right)^{2}\frac{d\Theta\left[\theta(z)\right]}{dz}\right\} \:.\nonumber \\
\label{AngMomDensity-Massive}
\end{align}
Now we can see that the two integral terms which violate the membrane
paradigm form have the same structure: the integrands are product
of effective masses of the vector fluctuations and non-trivial radial
flows of the axion profile. These two factors are the two sources
of the membrane paradigm violation. Near the boundary 
\[
m_{h}^{2}(z\rightarrow\infty)\rightarrow\frac{\Phi_{1}^{2}}{z^{4}}\:,\qquad m_{a}^{2}(z\rightarrow\infty)\rightarrow\frac{3\Gamma}{z^{3}}\:,
\]
the fluctuations become massless, as expected.\bigskip{}

\section{Tensor Mode Fluctuations and Hall Viscosity}

The Hall viscosity for gravitational Chern-Simons model with an axion
coupling has been computed in \cite{Saremi:2011ab,Chen:2011fs}, but
in a different form of the metric. Here, for completeness, we present
the derivation again, appropriate for the background ansatz (\ref{Background}).
The result we will derive here is also a generalization of the results
in \cite{Saremi:2011ab,Chen:2011fs}, since we have a generic Chern-Simons
coupling function $\Theta[\vartheta]$ in (\ref{GravCS_term}). In
this section, for computing viscosities, we only consider the homogeneous
case where all the fluctuations are independent of spatial coordinates
$x$ and $y$. In this case, the tensor mode fluctuations $h_{xy}$
and $h_{xx}-h_{yy}$ decouple from the rest.

\subsection{Tensor Mode EOMs and Solutions}

First we define: 
\begin{equation}
h_{xy}(t,z)=r(z)^{2}h_{e}(t,z)\:,\qquad\frac{1}{2}\left(h_{xx}(t,z)-h_{yy}(t,z)\right)=r(z)^{2}h_{o}(t,z)\:,
\end{equation}
where the subscripts $e$ and $o$ mean even and odd under parity
operation $x\leftrightarrow y$. We define the notations $\epsilon_{ij}$
($i=e,o)$ as following: $\epsilon_{eo}=-\epsilon_{oe}=1$, $\epsilon_{ee}=\epsilon_{oo}=0$.
The the linearized Einstein equations for these fluctuations (in momentum
space) are 
\begin{equation}
\frac{d}{dz}\left[r(z)^{2}F(z)\left(\frac{d}{dz}h_{i}(\omega,z)\right)\right]+\omega^{2}\frac{r(z)^{2}}{F(z)}h_{i}(\omega,z)=\Xi_{i}(\omega,z;\lambda)\:,
\end{equation}
where 
\begin{eqnarray}
\Xi_{i}(\omega,z;\lambda) & = & -\frac{i}{2}\omega\lambda\epsilon_{ij}\Bigg\{-2\frac{d}{dz}\left[r(z)^{2}F(z)\frac{d\bar{\Theta}(z)}{dz}\left(\frac{d}{dz}h_{j}(\omega,z)\right)\right]\\
 &  & \left[r(z)^{2}\frac{d}{dz}\left(\frac{dF(z)}{dz}\frac{d\bar{\Theta}(z)}{dz}\right)-2F(z)\frac{d}{dz}\left(r(z)\frac{dr(z)}{dz}\frac{d\bar{\Theta}(z)}{dz}\right)\right]h_{j}(\omega.z)\Bigg\}\nonumber \\
 &  & +i\omega^{3}\lambda\frac{r(z)^{2}}{F(z)}\frac{d\bar{\Theta}(z)}{dz}\epsilon_{ij}h_{j}(\omega,z)\nonumber 
\end{eqnarray}
and the repeated index $j$ is summed over $e$ and $o$. The incoming
wave solution is 
\begin{equation}
h_{i}(\omega,z)=\left(\frac{z-z_{H}}{z}\right)^{-i\frac{\omega}{4\pi T}}\left[\bar{h}_{i}+i\omega h_{i}^{(1)}(z)+O\left(\omega^{2}\right)\right]\:,\label{TensorSolution}
\end{equation}
where 
\begin{eqnarray}
h_{i}^{(1)}(z) & = & \bar{h}_{i}\left[\frac{1}{4\pi T}\ln\left(\frac{z-z_{H}}{z}\right)-r(z_{H})^{2}\int_{\infty}^{z}d\xi\frac{1}{r(\xi)^{2}F(\xi)}\right]\nonumber \\
 &  & +\lambda\epsilon_{ij}\bar{h}_{j}\Bigg\{2\pi Tr(z_{H})^{2}\bar{\Theta}'(z_{H})\int_{\infty}^{z}d\xi\frac{1}{r(\xi)^{2}F(\xi)}\\
 &  & -\frac{1}{2}\int_{\infty}^{z}d\xi\left[\frac{d}{d\xi}\ln\left(\frac{F(\xi)}{r(\xi)^{2}}\right)\right]\frac{d\bar{\Theta}(\xi)}{d\xi}\Bigg\}\:.\nonumber 
\end{eqnarray}
Because $\theta(z)$ is sourceless near the boundary: $\theta(z)\sim z^{-\Delta_{+}}$,
for a general $\Theta[\vartheta]=\vartheta^{n}$ ($n\geqslant1)$
near the boundary, the last line in the above equation goes to zero
faster than $O\left(z^{-3}\right)$ near the boundary, so only the
second line contributes to Hall viscosity.

\subsection{2-Point Functions and Viscosities}

The total second order on-shell action for the tensor mode, from (\ref{S2_bulk})
and the corresponding boundary terms, is 
\begin{eqnarray}
S^{(2)} & = & \frac{1}{4\kappa^{2}}\int_{z=\infty}d^{3}x\sum_{i,j=e,o}\Bigg\{\left[\frac{4}{L}r(z)^{2}\sqrt{F(z)}-\frac{d}{dz}\left(r(z)^{2}F(z)\right)\right]h_{i}^{2}-r(z)^{2}F(z)h_{i}\left(\frac{\partial}{\partial z}h_{i}\right)\nonumber \\
 &  & +\frac{\lambda}{2}\epsilon_{ij}\Bigg[r(z)^{4}\left(\frac{d}{dz}\frac{F(z)}{r(z)^{2}}\right)\frac{d\bar{\Theta}(z)}{dz}h_{i}\left(\frac{\partial}{\partial t}h_{j}\right)+2\frac{r(z)^{2}}{F(z)}\bar{\Theta}(z)\left(\frac{\partial^{2}}{\partial t^{2}}h_{i}\right)\left(\frac{\partial}{\partial t}h_{j}\right)\label{eq:S2_tensor}\\
 &  & +2r(z)\frac{dr(z)}{dz}F(z)\bar{\Theta}(z)\left(\frac{\partial}{\partial t}h_{i}\right)\left(\frac{\partial}{\partial z}h_{j}\right)-2r(z)^{3}F(z)\left(\frac{d}{dz}\frac{\bar{\Theta}(z)}{r(z)}\right)h_{i}\left(\frac{\partial^{2}}{\partial t\partial z}h_{j}\right)\Bigg]\Bigg\}\:.\nonumber 
\end{eqnarray}
Following the holographic prescriptions of \cite{Herzog:2002pc,Barnes:2010jp,Arnold:2011ja},
we obtain the 2-point functions in momentum space: 
\begin{eqnarray}
G_{\textrm{ra}}^{xy,xy}(\omega) & = & -\frac{\Gamma}{2\kappa^{2}L^{2}}-i\omega\frac{r(z_{H})^{2}}{2\kappa^{2}}+O\left(\omega^{2}\right)\:,\\
G_{\textrm{ra}}^{xx-yy,xy}(\omega) & = & -i\omega\frac{2\pi T\lambda}{\kappa^{2}}r(z_{H})^{2}\bar{\Theta}'(z_{H})+O\left(\omega^{2}\right)\:.
\end{eqnarray}
Comparing with Kubo formulae \cite{Paper1}:%
\footnote{The $\eta_{H}$ term differs by a sign from \cite{Paper1} because
our convention is $\epsilon_{txy}=1$ in 3-d flat Minkowskian space
(the boundary), which follows from $\epsilon_{txyz}=\sqrt{-g}$ in
4-d bulk space.%
} 
\begin{eqnarray}
G_{\textrm{ra}}^{xy,xy}(\omega) & = & p-i\omega\eta+O\left(\omega^{2}\right)\:,\\
G_{\textrm{ra}}^{xx-yy,xy}(\omega) & = & 2i\omega\eta_{H}+O\left(\omega^{2}\right)\:,
\end{eqnarray}
we get the shear viscosity 
\begin{eqnarray}
\frac{\eta}{s} & = & \frac{1}{4\pi}\:,
\end{eqnarray}
where the entropy density is $s=\frac{2\pi}{\kappa^{2}}r(z_{H})^{2}$,
and the Hall viscosity 
\begin{equation}
\eta_{H}=-\frac{\lambda}{4\kappa^{2}}\left\{ r(z)^{4}\left[\frac{d}{dz}\left(\frac{F(z)}{r(z)^{2}}\right)\right]\frac{d\Theta\left[\theta(z)\right]}{dz}\right\} \Bigg|_{z=z_{H}}=-\frac{\pi T\lambda}{\kappa^{2}}r(z_{H})^{2}\bar{\Theta}'(z_{H})\:,\label{HallViscosity}
\end{equation}
where $\bar{\Theta}'(z)\equiv\partial_{z}\Theta\left[\theta(z)\right]$.
The middle part in the above equation is from the first term in the
second line of (\ref{eq:S2_tensor}) and we have used $F(z)=4\pi T(z-z_{H})+O\left((z-z_{H})^{2}\right)$
to go to right hand side. The Hall viscosity has a simple form which
is expressed purely in terms of bulk quantities at the horizon. This
is a generalization of results in \cite{Saremi:2011ab,Chen:2011fs},
for a generic gravitational Chern-Simons term of form (\ref{GravCS_term}).
The gauge Chern-Simons term (\ref{GaugeCS_term}) has no contribution.
The reason is obvious: this term is totally independent of the metric;
since Hall viscosity is a response to the metric perturbation, it
is natural that the gauge Chern-Simons has no contribution. In contrary
to the membrane paradigm of the Hall viscosity, the angular momentum
density (\ref{AngMomDensity}) has a more complicated form: part of
it does have a membrane paradigm form while the rest is an non-trivial
bulk integral. This difference indicates that the physics behind these
two quantities are different (at least for the holographic Chern-Simons
models studied in this paper and the dual field theories they describe),
thus in general, we expect their ratio to have some non-trivial behavior.
This is difficult to study analytically. In the next section, we will
present the numeric results.

\bigskip{}

\section{Numeric Results of the Axion Condensate Phase}

In this section, we numerically study Hall viscosity, angular momentum
density and their ratio in terms of physical parameters such as temperature
$T$ and charge density $\rho$ (we will use Canonical ensemble in
this section, where $\rho$ is held fixed) in the axion condensate
phase of the holographic Chern-Simons model. In this phase, as the
temperature is lowered, the axion scalar $\vartheta$ develops a non-trivial
profile in the bulk and thus breaks the parity spontaneously. The
order parameter corresponds to the expectation value $\langle\mathcal{O}\rangle$
of the operator $\mathcal{O}$ in the field theory that is dual to
the scalar $\vartheta$. We will first work in the probe limit, and
then include full back-reactions.

From (\ref{AngMomDensity}) we see that angular momentum density receive
contributions from both gauge and gravitational Chern-Simons terms.
However, from (\ref{HallViscosity}), Hall viscosity is only determined
by the gravitational Chern-Simons term. In general, the two Chern-Simons
coupling functions $\lambda_{A}\Theta_{A}[\vartheta]$ and $\lambda\Theta[\vartheta]$
can be different and unrelated. To make our analysis simple, from
now on in most of this section, we will focus only on the gravitational
Chern-Simons term, and turn off the gauge Chern-Simons term $\lambda_{A}=0$.
Only at the end of this section will we include the numeric result
for the angular momentum density from the gauge Chern-Simons model.

Before starting the numeric analysis, we would like to first discuss
the choice of the general function $\Theta[\vartheta]$ in (\ref{GravCS_term}).
There is no unique choice for its form from the phenomenological model
we write down here. When $\Theta[\vartheta]=\textrm{constant}$, the
gravitational Chern-Simons term (\ref{GravCS_term}) is a boundary
term because the Pontryagin density is a total derivative. This is
not the case we are interested here, because we want this term to
be dynamical, at least to have a non-trivial $z$-profile to generate
non-vanishing Hall viscosity. In \cite{Saremi:2011ab,Chen:2011fs,Chen:2012ti},
the authors chose $\Theta[\vartheta]=\vartheta$, which results in
the near-critical behavior of Hall viscosity to be $\eta_{H}\sim(T_{c}-T)^{1/2}$,
because the Hall viscosity is linear to the condensate $\theta\sim\langle\mathcal{O}\rangle\sim(T_{c}-T)^{1/2}$.
Another form, $\Theta[\vartheta]=\vartheta^{2}$, is also interesting,
because from (\ref{AngMomDensity}) and (\ref{HallViscosity}), both
Hall viscosity and angular momentum density are quadratic in order
parameter $\langle\mathcal{O}\rangle$ now, so near critical regime
they will scale as $T_{c}-T$, instead of $(T_{c}-T)^{1/2}$. This
is in agreement with condensed matter theory arguments such as in
$p_{x}+ip_{y}$ paired states of BCS theory: since both Hall viscosity
and angular momentum density have dimension $1/[\textrm{length}]^{2}$,
by dimensional analysis, they are proportional to square of the order
parameter, thus scale as $T_{c}-T$ near critical regime. Of course
$\Theta[\vartheta]$ can take other forms in general. For $\Theta[\vartheta]=\vartheta^{n}$,
Hall viscosity and angular momentum density will scale as $(T_{c}-T)^{n/2}$
near critical regime and $\eta_{H}/\ell$ will acquire a factor of
$n$. When $\Theta[\vartheta]$ contains multiple terms of $\vartheta$
with different powers, the lowest power will dominate the near-critical
behavior and the highest power the low temperature behavior. In this
section we choose 
\begin{equation}
\Theta[\vartheta]=\vartheta^{2}
\end{equation}
so as to reproduce the $T_{c}-T$ scaling near the critical regime.

\subsection{Gravitational Chern-Simons Model: The Probe Limit}

\begin{figure}[t]
\begin{centering}
\includegraphics[width=7cm]{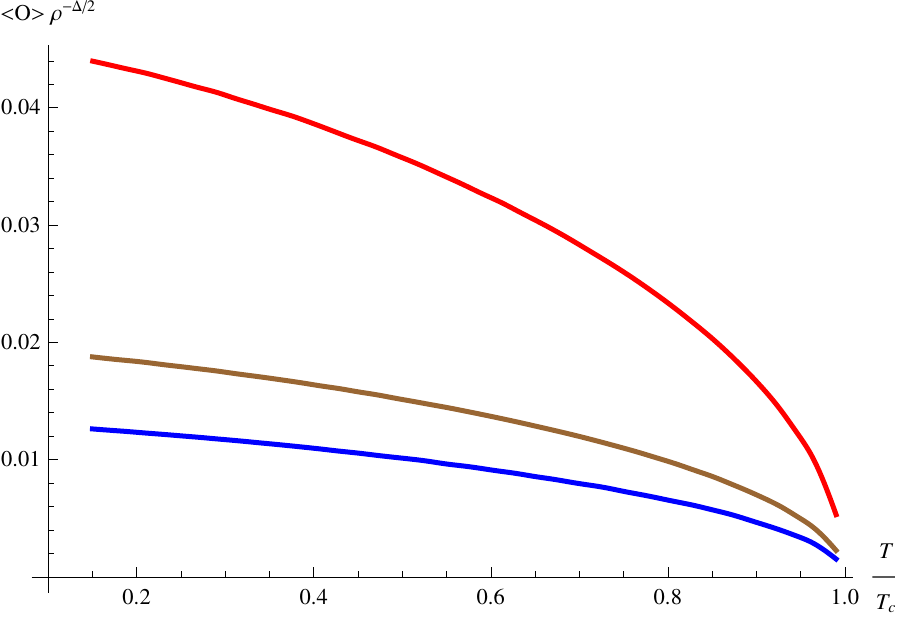}\qquad{} \includegraphics[width=7cm]{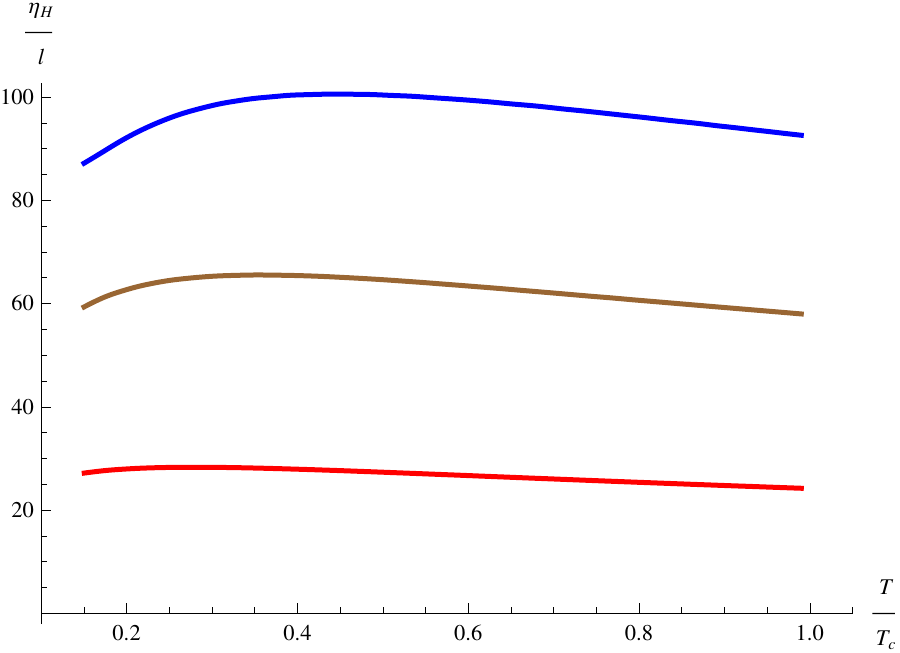}
\par\end{centering}

\begin{centering}
\includegraphics[width=7cm]{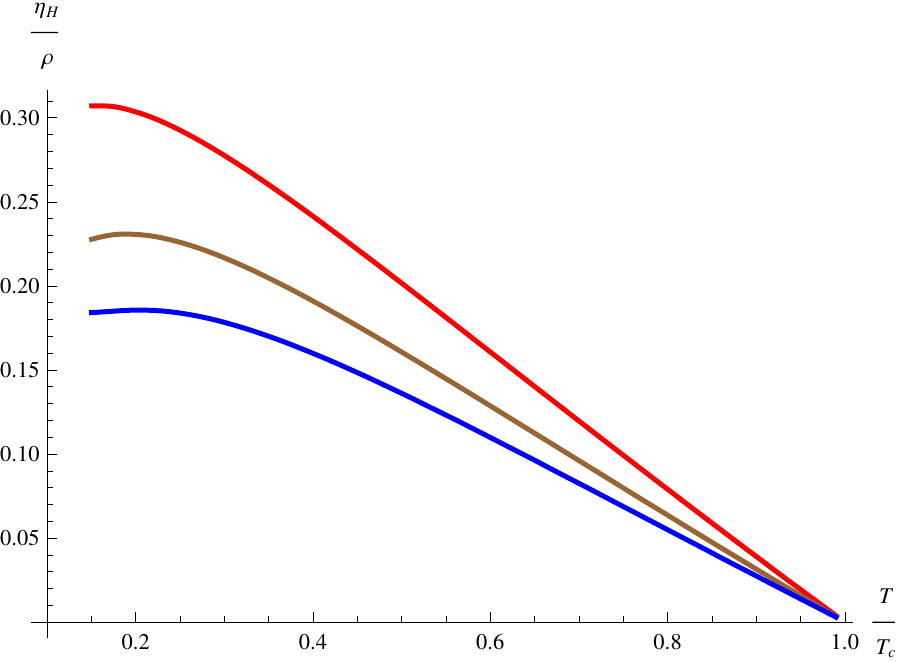}\qquad{} \includegraphics[width=7cm]{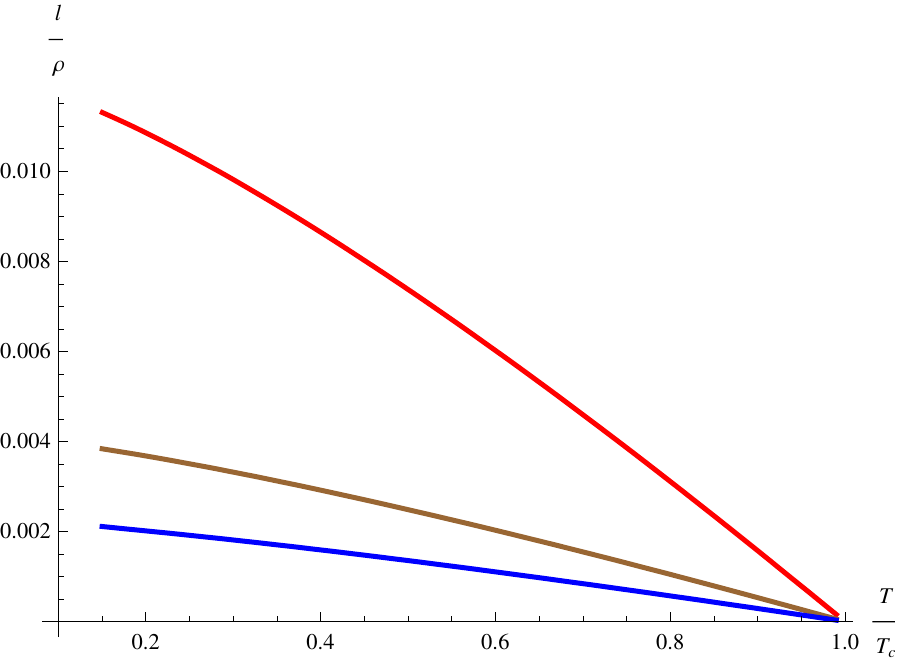}
\par\end{centering}

\caption{Condensate $\langle\mathcal{O}\rangle$ (upper left), Hall viscosity
$\eta_{H}$ (lower left), angular momentum density $\ell$ (lower
right) and their ratio $\eta_{H}/\ell$ (upper right) as functions
of $T/T_{c}$ in the probe limit, with $c_{4}L^{2}=0.5$. The red,
brown and blue lines correspond to $m^{2}L^{2}=-2.2$, $-2.1$ and
$-2.05$, respectively. We have set $L=\lambda=\kappa=1$, $\lambda_{A}=0$.}
\label{Fig:ProbeLimit}
\end{figure}

In this subsection we study the probe limit of the bulk theory, where
the scalar field $\vartheta$ does not back-react on the metric and
Maxwell field. This limit has been employed in \cite{Chen:2011fs,Chen:2012ti}.
The background now is the AdS-Reissner-Nordström black hole:
\begin{equation}
\left\{ \begin{aligned} & r(z)=\frac{z}{L}\\
 & F(z)=\left(\frac{z}{L}\right)^{2}-\left(1+\frac{Q^{2}}{4}\right)\frac{z_{H}^{3}}{L^{2}z}+\frac{Q^{2}z_{H}^{4}}{4L^{2}z^{2}}\\
 & \Phi(z)=Q\frac{z_{H}}{L}\left(1-\frac{z_{H}}{z}\right)
\end{aligned}
\:,\right.\label{AdSRN4}
\end{equation}
where $Q$ is the dimensionless charge. The temperature is 
\begin{equation}
T=\frac{3z_{H}}{4\pi L^{2}}\left(1-\frac{Q^{2}}{12}\right)\:,
\end{equation}
and to make the temperature non-negative, the charge has to satisfy
$Q^{2}\leqslant12$. The near-boundary behavior of the scalar field
is 
\begin{equation}
\theta(z)\rightarrow\frac{\langle\mathcal{O}\rangle}{z^{\Delta_{+}}}\:,
\end{equation}
and that of the electric potential is 
\begin{equation}
\Phi(z)=\mu-\frac{\rho}{z}+O\left(\frac{1}{z^{2}}\right)\:,
\end{equation}
where $\langle\mathcal{O}\rangle$ is the condensate, $\mu$ the chemical
potential and $\rho$ the charge density (up to some factors of $\kappa$
and $L$). Mass $m$ is related to the conformal dimension $\Delta_{+}$
of the condensate operator $\mathcal{O}$ by (\ref{ConformalDimension}).
It has been shown in \cite{Chen:2011fs} that in this setup the black
hole can develop a scalar hair only at very low temperature, where
it is near extremal and the extremality factor $1-Q^{2}/12$ is close
to zero. Figure (\ref{Fig:ProbeLimit}) shows numeric results for
$c_{4}L^{2}=0.5$ with various values of $m$. It is interesting to
notice that despite the seemingly different analytic expressions for
Hall viscosity and angular momentum density, their ratio $\eta_{H}/\ell$
remains more or less unchanged for a vast range of temperature until
one reaches the very low temperature regime. However, the value of
the ratio is typically a huge number depending on the mass $m$ (or
conformal dimension $\Delta_{+}$) of the scalar condensate, and is
far away from the $1/2$ value found in condensed matter literature.
We will see that this feature remains when full back-reactions are
included. 

The stability for a charged black hole with a neutral scalar condensation
was discussed in \cite{Iqbal:2010eh} as well as in \cite{Chen:2011fs,Chen:2012ti}.
In this paper when talking about neutral scalar condensation, we always
focus on a narrow window around $m^{2}L^{2}=-2$, which is with in
the range discussed in these references and the black hole can develop
a neutral scalar hair which condensate near the horizon.

\subsection{Gravitational Chern-Simons Model: Including Back-reactions}

The probe limit usually works well in high temperature when the black
hole is far from extremal, and the condensate is small and the back-reactions
are weak. However, in the probe limit of the previous subsection,
numerics shows that the scalar can only condensate when the black
hole is near-extremal. But in this case the back-reactions play a
very important role, thus the probe limit assumption may not be consistent.
Particularly, Hall viscosity is solely expressed in terms of quantities
near the horizon and so is part of the angular momentum density, thus
the accuracy of the numeric solutions near the horizon matters a lot.
Even in high temperature, when the back-reactions are negligible near
the boundary and not strong in most part of the bulk, they are still
very important near the horizon. For example, $r(z)$ has significant
deviation from its probe limit form near the horizon. To improve the
accuracy of the numeric results, in this subsection we will take back-reactions
into full account.
\begin{figure}[t]
\begin{centering}
\includegraphics[width=7cm]{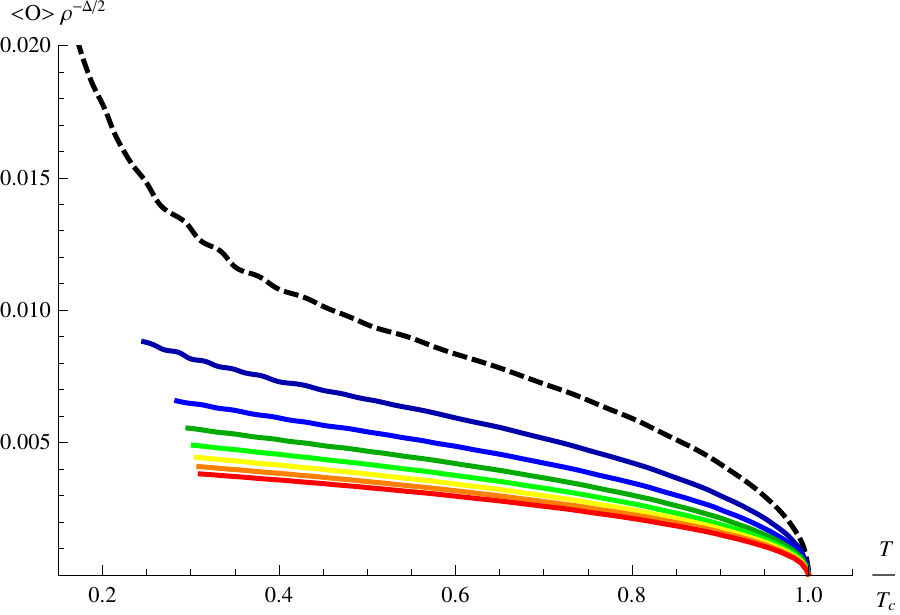}\qquad{}
\includegraphics[width=7cm]{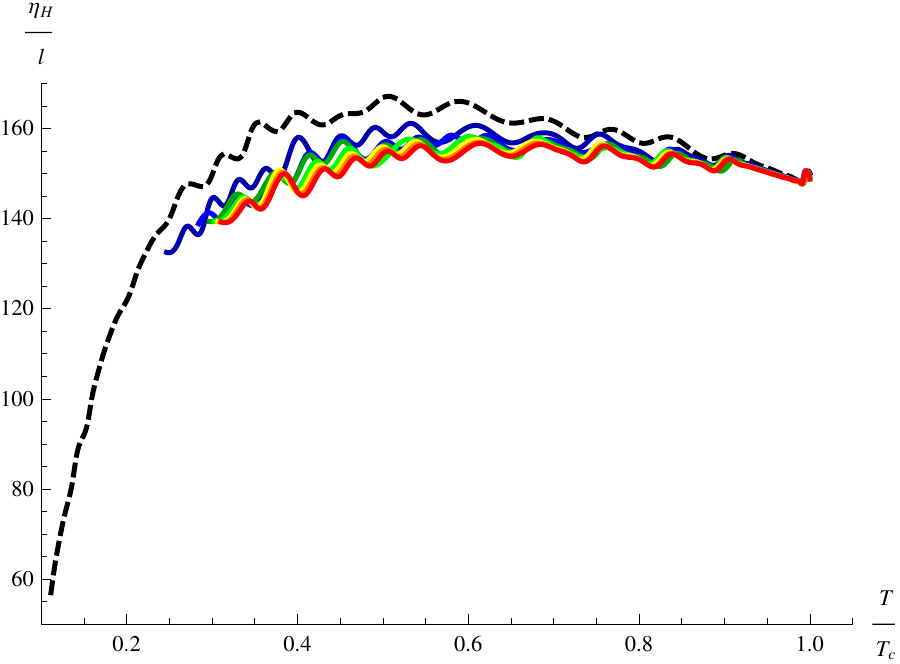}
\par\end{centering}

\begin{centering}
\includegraphics[width=7cm]{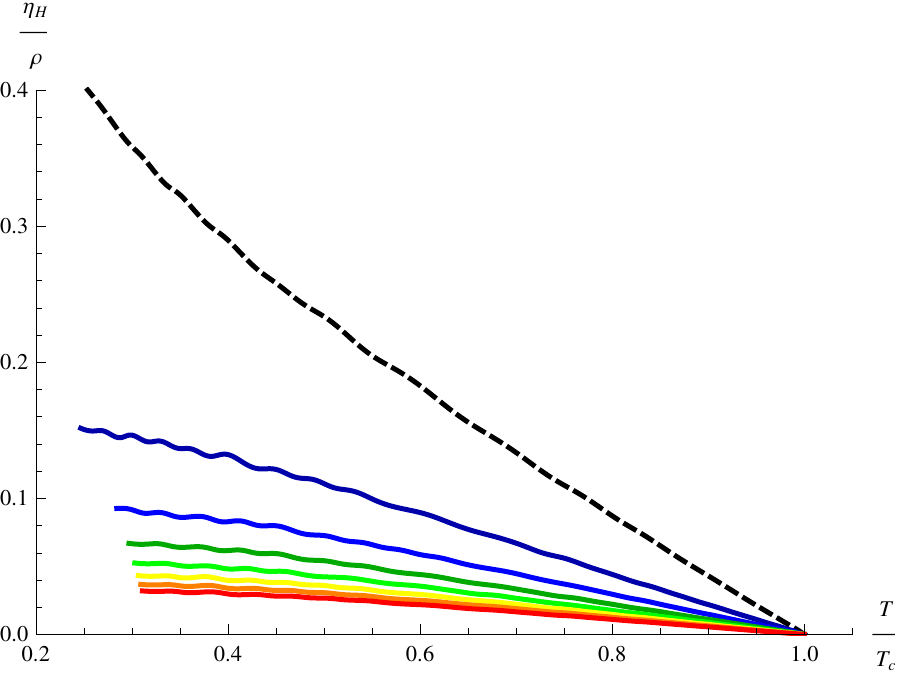}\qquad{}
\includegraphics[width=7cm]{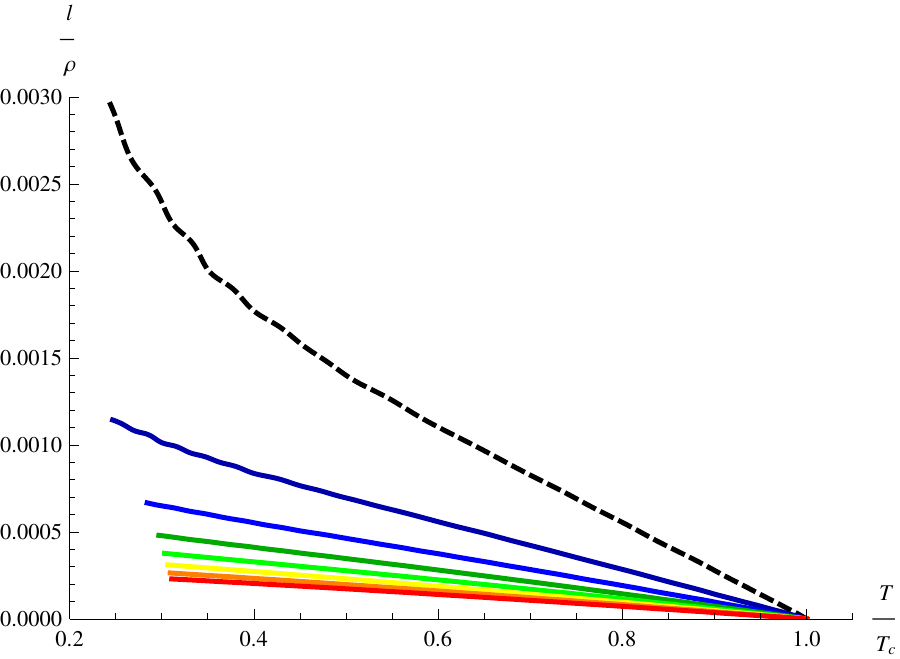}
\par\end{centering}

\caption{Condensate $\langle\mathcal{O}\rangle$ (upper left), Hall viscosity
$\eta_{H}$ (lower left), angular momentum density $\ell$ (lower
right) and their ratio $\eta_{H}/\ell$ (upper right) as functions
of $T/T_{c}$, with full back-reactions included. Here $m^{2}L^{2}=-2$,
corresponding to conformal dimension $\Delta_{+}=2$. The black dashed
line has $c_{4}L^{2}=0$, with an increment of $0.25$ for each adjacent
line toward the red one, which has $c_{4}L^{2}=1.75$. We have set
$L=\lambda=\kappa=1$, $\lambda_{A}=0$.}
\label{Fig:FullBackreaction}
\end{figure}

Figure (\ref{Fig:FullBackreaction}) shows the numeric results when
full back-reactions are included. The mass is chosen to be $m^{2}L^{2}=-2$,
corresponding to conformal dimension $\Delta_{+}=2$. The black dashed
line is the linear case with $c_{4}=0$. The colored lines show non-linear
effect, with bigger non-linear coefficient $c_{4}$ toward the red
end. The non-linearity decreases the values of the condensate, Hall
viscosity and angular momentum density. Numerically we find that below
certain low temperature ($<0.3T_{c}$) it is hard to find a condensate
solution when $c_{4}>0$: that is the reason why all the colored curves
terminate at some low temperature. 

The ratio between Hall viscosity and angular momentum density $\eta_{H}/\ell$
remains more or less unchanged at high temperature, same as in the
probe limit. It only starts to drop off dramatically once gets to
low temperature regime where $T<0.3T_{c}$. It is not clear to us
whether there is a physical origin or interpretation of the wiggles
in the plot. Non-linearity has almost no effect near the critical
temperature, because here the condensate is close to zero and the
non-linear term is of higher order. It will only show up when the
temperature is lowered and the condensate becomes large enough such
that the non-linear term is comparable to the other terms. The non-linearity
does decrease the ratio, however, its effect to the ratio is much
weaker compared to that to Hall viscosity and angular momentum density
individually. The numeric plot suggests that the non-linear effect
on the $\eta_{H}/\ell$ ratio is at its strongest at the mid-temperature
regime where $T\approx0.5T_{c}$. As the temperature is lowered further,
the non-linear effect on the ratio shows a trend to become weaker
as the colorful lines go closer. It is interesting to notice that
as the temperature drops below $T\approx0.2T_{c}$, the ratio decrease
dramatically towards zero. Due to the difficulties of numeric calculation
for extremely low temperature regime, we can only work out the dashed
line ($c_{4}=0$) below $T\approx0.25T_{c}$ and can not go beyond
$T\lesssim0.1T_{c}$. It will be interesting to see whether at zero
temperature the ratio approaches some non-zero fixed value, for example,
$1/2$ widely found in the study of field theory and condensed matter
systems \cite{Read:2008rn,Read:2011,Hoyos:2011ez,Nicolis:2011ey,Son:2013rqa}.
In this regime, the near horizon geometry may have different scalings
than the $\mathrm{AdS}_{2}\times\mathbb{R}^{2}$ of extremal black
holes, such as a $\mathrm{AdS}_{2}$ with a different radius, and
across a critical conformal dimension $m_{c}^{2}L^{2}=-3/2$ when
$T_{c}$ reaches zero, the system may undergo a Berezinskii-Kosterlitz-Thouless
type phase transition with an exponentially generated scaling \cite{Iqbal:2010eh}.
We will leave the study on zero temperature regime to the future.
\begin{figure}[t]
\begin{centering}
\includegraphics[width=7cm]{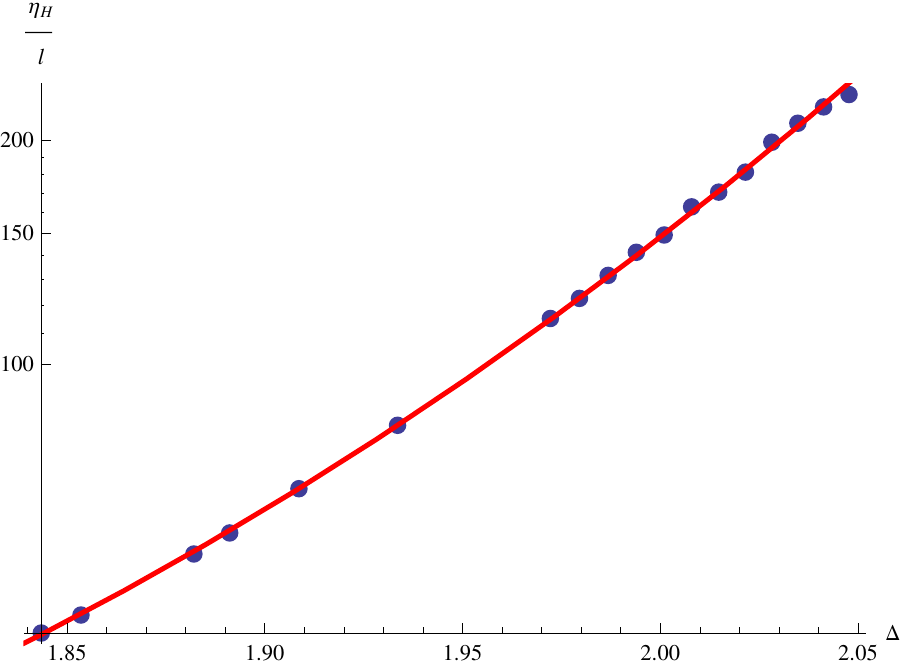}
\par\end{centering}

\caption{Hall viscosity to angular momentum density ratio $\eta_{H}/\ell$
near the critical regime $T\rightarrow T_{c}$ as a function of conformal
dimension $\Delta_{+}$. $c_{4}L^{2}=\frac{1}{120}$. The dots are
the numeric data and the red line is the fitting of (\ref{RatioNC_mass2}).
The vertical axis is shown in logarithmic scale.}
\label{Fig:RatioNearCritical}
\end{figure}

The ratio also depends on the mass $m$ (the conformal dimension $\Delta_{+}$)
as in the probe limit. Since the non-linearity does not play an important
role, to separate this effect, we can just study the ratio's dependence
on the mass at the critical temperature. Figure (\ref{Fig:RatioNearCritical})
shows the near-critical $\eta_{H}/\ell$ ratio as a function of conformal
dimension, with a fitting of the following form: 
\begin{equation}
\frac{\eta_{H}}{\ell}\Bigg|_{T\rightarrow T_{c}}=e^{-26.8\sqrt{\Delta_{+}(3-\Delta_{+})}+42.9}\:,\label{RatioNC_mass2}
\end{equation}
i.e. this near-critical ratio depends on mass $m$ exponentially.

\subsection{Gauge Chern-Simons Model: Angular Momentum Density}

\begin{figure}[t]
\begin{centering}
\includegraphics[width=8cm]{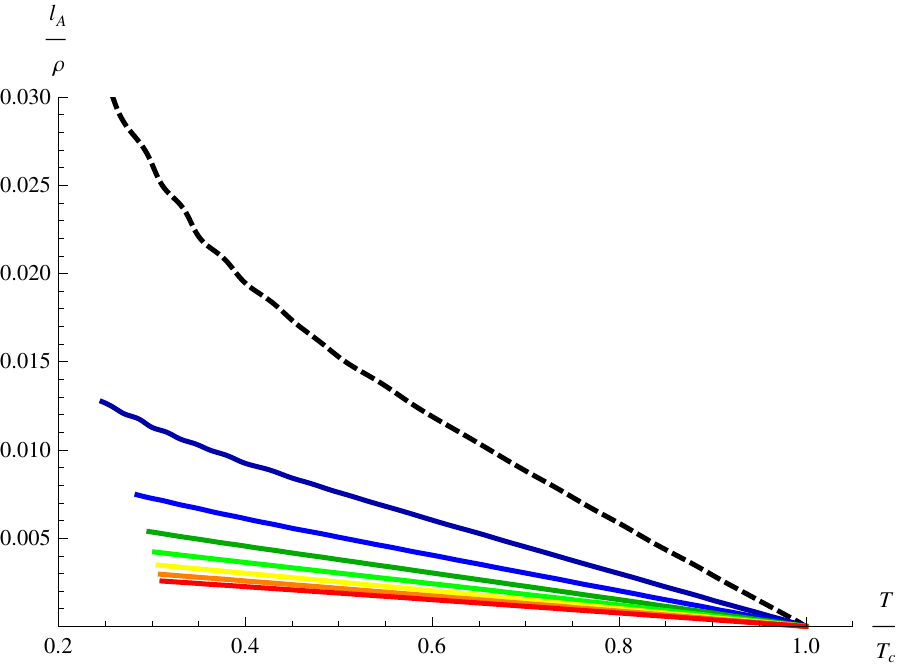}
\par\end{centering}

\caption{Angular momentum density $\ell$ as a function of $T/T_{c}$, with
full back-reactions included. Here $m^{2}L^{2}=-2$, corresponding
to conformal dimension $\Delta_{+}=2$. The black dashed line has
$c_{4}L^{2}=0$, with an increment of $0.25$ for each adjacent line
toward the red one, which has $c_{4}L^{2}=1.75$. We have set $L=\lambda_{A}=\kappa=1$,
$\lambda=0$.}
\label{Fig:GaugeCherSimons}
\end{figure}

At the end of this section, in Figure (\ref{Fig:GaugeCherSimons}),
we present the numeric result for the angular momentum density due
to the gauge Chern-Simons term, i.e. the first line in (\ref{AngMomDensity}).
We turn off the gravitational Chern-Simons coupling: $\lambda=0$
and choose $\Theta_{A}[\vartheta]=\vartheta^{2}$ as well. Qualitatively
the plot is very similar to that of the gravitational Chern-Simons
model. 

\bigskip{}

\emph{Note added}: After the original version of this paper, \cite{Liu:2014gto}
appeared, where the angular momentum density, Hall viscosity and their
ratio are studied analytically or numerically in a few classes of
holographic models. The general form of the action they start with
is the same as ours, with choices of different forms of the scalar
potential $V\left[\vartheta\right]$. In the first class of models
studied there, the non-normalizable mode of the scalar $\vartheta$
is turned on. This corresponds to turn on the source $\theta_{0}$
here. In principle, the sourceless case we study in this section can
be viewed as a limiting case as $\theta_{0}\rightarrow0$, given all
other settings are the same in the $\theta_{0}\neq0$ case. Unfortunately
this does not happen when we try to compare our results with those
of \cite{Liu:2014gto}, because other settings are not quite the same
between ours and theirs. In our sourceless case, if the charge density
$\mu$ and chemical potential $\rho$ are turned off, the condensate
$\theta(z)\neq0$ can not form and both $\eta_{H}$ and $\ell$ vanishes
identically. In this case their ratio is simply not well defined (in
other words, $0/0$ can be anything). Sections III.B and III.C of
\cite{Liu:2014gto} consider cases without $\mu$ or $\rho$, but
with source on. Although their $\eta_{H}/\ell$ does have a good limit
as $\theta_{0}\rightarrow0$, this is not comparable to ours, and
we suspect that the order of two limits $\theta_{0}\rightarrow0$
and $\mu,\rho\rightarrow0$ may not be exchangeable, since they correspond
to approaching the non-analytic point of $0/0$ of $\eta_{H}/\ell$
from different directions in the parameter space. For this same reason,
results in section III.D and E are not immediately comparable either.
Superficially, the case closest to what we study is section III.D,
but only part of the angular momentum density, the non-integral part
(called $\mathcal{J}_{\mathrm{horizon}}$ there), is presented; the
non-trivial integral part is omitted, thus a direct comparison is
not available. In section IV of \cite{Liu:2014gto} the sourceless
case is studied directly, this is the same as what we have studied
in this section, where $\theta_{0}=0$ is imposed from the very beginning.
But they turn on a non-trivial dilaton coupling $e^{-\alpha\vartheta}$
for the Maxwell term in the action, where we set it to unity. In their
numeric computation, $\alpha$ is no less than $0.5$, thus a limiting
case when $\alpha\rightarrow0$ is not available for comparison. When
$\alpha$ is finite, the dilaton coupling facilitates the formation
of condensate: from their results we can see the non-trivial profile
of $\vartheta$ can form at any temperature, especially at very high
temperature; whereas for ours with $\alpha=0$, it can only form below
certain critical temperature $T_{c}$. Again this implies the high
temperature (or small $\mu$) limit and $\alpha\rightarrow0$ limit
may not be exchangeable when dilaton coupling is included. In summary,
none of the cases studied in \cite{Liu:2014gto} encloses ours as
a simple limiting case and they are complementary to what is studied
in this section.

\bigskip{}

\section{Conclusions and Comments}

We have shown that holographic models with gauge and/or gravitational
Chern-Simons terms have a non-vanishing angular momentum density when
parity is broken by the scalar coupled to the Chern-Simons terms.
Unlike Hall viscosity, the angular momentum density (\ref{AngMomDensity})
does not have a membrane paradigm form: it is not solely determined
by the near-horizon behavior of the background fields; part of it
is an integral over the whole bulk regime outside the horizon, which
suggests that the UV and IR degrees of freedom that are responsible
for the generation of angular momentum density interact non-trivially
with each other and do not decouple. These results are in agreement
with those obtained in \cite{Liu:2013cha}. The effect of this angular
momentum density is to accumulate momentum at the 1-dimensional spatial
boundary of the 2+1-dimensional system, inducing an edge current of
momentum whose strength is proportional to the angular momentum density,
as shown in \cite{Paper1} and \cite{Liu:2013cha}. 

We have presented numeric results for the axion condensate phase of
the gravitational Chern-Simons model when the scalar condensates,
both at the probe limit and with back-reactions fully included. Both
Hall viscosity and angular momentum density are monotonically decreasing
functions of temperature. Non-linearity of the scalar potential $V[\vartheta]$
plays little role in asymptotic behaviors near the critical regime,
but decreases both Hall viscosity and angular momentum density below
the critical temperature. 

The Hall viscosity to angular momentum density ratio obtained numerically
from the gravitational Chern-Simons term alone is not exactly a constant,
but it does remain more or less unchanged for a vast range of temperature,
except at the very low temperature regime. On the other hand, the
ratio obtained from non-holographic approaches \cite{Read:2008rn,Read:2011,Hoyos:2011ez,Nicolis:2011ey,Son:2013rqa}
is always $1/2$. In fact, the apparently different forms of (\ref{HallViscosity})
and (\ref{AngMomDensity}), and the facts that the former involves
only gravitational Chern-Simons term and tensor mode metric fluctuations
while the latter involves both gauge and gravitational Chern-Simons
terms and only vector mode fluctuations, already suggest that the
physical mechanisms of generating Hall viscosity and angular momentum
density in holographic Chern-Simons models and the dual field theories
they describe are quite different. Thus in general, a simple relationship
between them would not be expected from these theories. How to understand
the universal relation obtained from field theory and condensed matter
theories and the non-universal results from holographic Chern-Simons
models here (and what role the gauge Chern-Simons term plays regarding
the relationship between Hall viscosity and angular momentum density)
is still open questions to be answered in the future.

In this paper, we have obtained a general analytic formula (\ref{AngMomDensity})
for angular momentum density. But numerically we have only studied
the axion condensate phase where the AdS-Reissner-Nordström black
hole develops a neutral scalar hair. At zero or very low temperature,
the system may flow to different infrared geometries with different
scalings \cite{Iqbal:2010eh,Gouteraux:2012yr}. How the angular momentum
density, Hall viscosity and their ratio behave in these different
IR fixed points, is another interesting question that can be studied
in the future.

\bigskip{}

\section*{Acknowledgments \textmd{\normalsize{\addcontentsline{toc}{section}{Acknowledgments}}}}

The author is very grateful to Dam Thanh Son for inspiring discussions
throughout the progress of this work, and thanks Nien-En Lee, Hong
Liu, Hirosi Ooguri, Nicholas Read, Bogdan Stoica and Paul Wiegmann
for informative comments and useful email communications. The author
thanks particularly the JHEP referee for very thoughtful comments
and suggestions. This work is supported by DOE grant DE-FG02-90ER-40560
and NSF DMS-1206648.

\bigskip{}
\bigskip{}

\end{document}